\documentclass{article}
\usepackage{graphicx} 
\usepackage{float}
\usepackage{biblatex} 
\usepackage{hyperref}
\usepackage{caption}
\usepackage{booktabs}
\usepackage{multirow} 
\usepackage{tabularray}  
\usepackage{tabularx}
\usepackage{longtable}
\usepackage{subcaption}
\usepackage{algorithm}
\usepackage{multirow}
\usepackage{algpseudocode}
\usepackage{amsmath}
\usepackage{amsfonts}
\usepackage{setspace}
\usepackage{siunitx}
\usepackage{indentfirst}

\usepackage{helvet}                        

\title{Multimodality Stacking with Blockwise missing values and application to the PIONeeR biomarkers study for prediction of resistance to immunotherapy}
\author{Mohamed Boussena$^1$  \and Florence Monville$^2$ \and Jacques Fieschi-Meric$^2$ \and Frederic Vely$^3$ \and Pierre Milpied$^3$  \and Julien Mazieres$^4$  \and Maurice Perol$^5$  \and Eric Vivier$^6$ \and Laurent Greillier$^1$$^3$ \and Fabrice Barlesi$^7$ \and Sebastien Benzekry$^1$ }

\date{%
    $^1$Inria – Inserm team COMPO, COMPutational pharmacology and clinical Oncology, Centre Inria Sophia Antipolis -
Méditerranée, Centre de Recherches en Cancérologie de Marseille, Inserm U1068, CNRS UMR7258, Institut
Paoli-Calmettes, Pharmacy faculty, Aix-Marseille University\\%
    $^2$Veracyte SAS, Marseille, France\\%
    $^3$Assistance Publique-Hôpitaux de Marseille (APHM), Marseille, France\\%
    $^4$ Toulouse University Hospital, Toulouse, France\\%
    $^5$Centre Leon Berard, Lyon, France\\%
    $^6$Innate Pharma, Marseille, France\\%
    $^7$Université Paris Saclay, Gustave Roussy, Inserm, Prédicteurs Moléculaires et nouvelles cibles en oncologie (U981), F-94805, Villejuif, France\\%
}

\begin{document}
\maketitle


\newpage 

\section*{Abstract}

Integrating multimodal datasets in clinical oncology is frequently hindered by high dimensionality and blockwise missingness, where entire data sources are unavailable for specific patient subsets. Standard survival models often struggle with these gaps, leading to biased results or patient exclusion.

We introduce Multimodality Stacking with Blockwise missing values (MSB), a late-fusion survival framework that fits base learners independently on each data source and aggregates their out-of-fold risk scores in a cross-validated stacking meta-learner. Because each source is represented by its risk scores rather than its raw features, an absent block reduces to a few missing features instead of hundreds of missing covariates.

Benchmarked against 12 multi-omics survival models across 21 public TCGA cohorts (SurvBoard), MSB ranked first by median time-dependent concordance and, when fitted withmissing-modalities, first on both concordance and integrated Brier score. 

We illustrate MSB applied on the PIONeeR (Precision Immuno-Oncology for advanced Non-Small Cell Lung Cancer Patients with PD1(L1) ICI Resistance) biomarkers study (437 patients, 378 biomarkers across eight heterogeneous sources) to predict progression-free survival. MSB identified routine markers — clinical features, medical biology, and PD-L1 expression — as the dominant predictors of progression-free survival, matching or exceeding invasive tumor-derived profiling; missing-block indicators contributed negligibly.

On PIONeeR, MSB significantly outperformed the unified baselines on discrimination (Wilcoxon signed-rank, Holm-corrected: $+14.4\%$ linear, $+5.0\%$ forest, $+1.8\%$ boosting; all $p<0.05$), the sole exception being gradient boosting, where the difference was not significant.

Implementation of the MSB framework is available at \url{https://github.com/MohamedBoussena/MSB}.

\newpage 
\section{Introduction}

In many biomedical applications, three major issues frequently occur: (1) high dimensionality ($n \sim p$ or $n \ll p$), (2) a substantial proportion of missing data and (3) high multicollinearity among features within the same data source. While conventional feature selection and ensemble-based algorithms aim to mitigate the `curse of dimensionality' \cite{salerno_high-dimensional_2023}, they exhibit significant statistical instability in small cohorts, especially among correlated predictors. These issues are compounded with missing values as standard imputation techniques lose fidelity as dimensionality increases. Given that high dimensionality in clinical datasets often arises from the integration of heterogeneous data sources, we hypothesize that addressing the problem at the source level — rather than the feature level — provides a more robust framework for survival modeling.

This source-level viewpoint is consistent with an active area of machine learning research: multimodal learning, closely related to multi-kernel, mixture-of-experts, and multiview learning \cite{NIPS1990_432aca3a,xu2013surveymultiviewlearning,krones_review_2025, gonen_multiple_2011}. Although it is frequently used to integrate heterogeneous data types (such as tabular data and images), multimodal learning also provides a principled framework for managing tabular data originating from multiple sources. Three main classes of multimodal learning approaches can be distinguished: early fusion, intermediate fusion, and late fusion \cite{krones_review_2025}. 

Early fusion refers to methods that combine the different modalities before the actual learning step. In practice, this typically amounts to a direct concatenation of features from all modalities into a single dataset, possibly after some preprocessing or transformation. This represents the most straightforward and, in many applications, the default strategy for multimodal tabular data. Intermediate fusion, by contrast, covers approaches in which fusion and learning take place jointly—for example, by learning modality-specific weights or by transforming the data on the fly during training. Late fusion describes approaches that first train individual models on each modality and then integrate their predictions using a higher-level model or a rule-based aggregation scheme. This categorization can be broadened to include “hybrid fusion” strategies \cite{zhao_deep_2024}, which denote any method that blends elements of the three core fusion types. 

In bioinformatics, this multimodal structure typically appears as multi-omics data. Chen et al. \cite{chen_applications_2023} provide an overview of the different components of multi-omics, which can be summarized as measurements of distinct molecular layers—genomic, transcriptomic, epigenomic, proteomic—obtained from the same sample. Each omic is generated by a different experimental assay and therefore constitutes a natural source, making multi-omics a prototypical example of source-level heterogeneity. As profiling technologies have become less expensive and interest in elucidating biological mechanisms has intensified, such datasets have grown both more abundant and more high-dimensional, exacerbating issues of extreme dimensionality, blockwise missing data, and strong within-source multicollinearity. 

While multimodal integration of imaging, omics, and tabular data offers significant potential for biomarker discovery \cite{steyaert_multimodal_2023}, clinical oncology datasets are frequently constrained by small sample sizes and high rates of missing values. Regardless of the fusion strategy considered, current multimodal approaches commonly address blockwise missingness through listwise deletion — the exclusion of patients lacking complete data across all modalities \cite{keyl_multimodal_2022, captier_integration_2025}. Beyond the substantial reduction in effective sample size this entails, such exclusions are rarely justified by the underlying missingness mechanism and have been shown to introduce systematic bias in predictive models \cite{van_buuren_flexible_2018}.

Missing data in multimodal learning is a well-studied issue, and numerous methods have been proposed in both traditional machine learning and deep learning settings~\cite{wu_deep_2024,baena-miret_framework_2024,flores_missing_2023}. For smaller sample sizes, Xiang et al.~\cite{xiang_bi-level_2014} introduced a late-fusion model in which an intermediate model is trained per modality, when present, and the resulting scores are imputed for missing modalities; a meta-model is then trained on this reduced representation. Building on this idea, van Loon et al.~\cite{van_loon_imputation_2024} proposed Stacked Penalized Logistic Regression (StaPLR), which trains an intermediate classifier per modality; performing feature selection at the modality level ; and imputes the resulting scores for downstream classification.

A second line of work addresses missing modalities by bypassing their influence rather than imputing them. Captier et al.~\cite{captier_integration_2025}, investigating the same clinical problem as we do—first-line immunotherapy outcomes in metastatic NSCLC—compared a weighted-average late fusion with an attention-based late fusion, each handling absent modalities by assigning them a weight of zero. HEALNet~\cite{hemker_healnet_2024} adopts a similar strategy in a survival analysis context, using cross-modal attention to dynamically weight the observed modalities and modeling a discrete hazard over time-binned intervals; however, it is a deep learning approach specialized for imaging that treats tabular covariates on a per-feature basis rather than as a single, cohesive modality.

Blockwise missingness and right-censored survival are therefore each well-addressed in isolation, but jointly only by deep architectures whose sample-size requirements are ill-matched to single-cohort clinical studies. MSB occupies this gap: it retains the score-level representation of the StaPLR lineage, extends it from classification to survival, and requires no modality-weighting scheme to be learned.

Here, we departed from a nationwide clinical dataset collected to address a current problem in immuno-oncology: to predict the onset of resistance to immune-checkpoint inhibitors (ICI) in advanced non-small cell lung cancer (NSCLC). Indeed, long-term remissions are only achieved in 20-30\% of such patients \cite{borghaei_five-year_2021,garassino_pembrolizumab_2023,gandhi_pembrolizumab_2018}, making lung cancer the leading cause of cancer-related deaths \cite{bray_global_2024}. A French national consortium grouping 17 clinical centers and 6 biological and industrial partners was constituted within the PIONeeR (Precision Immuno-Oncology for advanced Non-Small Cell Lung Cancer Patients with PD1(L1) ICI Resistance) clinical study (2018-2024) \cite{barlesi_integrative_2026}. The biological markers from either the peripheral blood or tumor tissue collected pre-treatment yielded a dataset with 378 biomarkers in 437 patients. Because each partner processed samples to generate specific biomarker types, the data exhibited both a multi-modal structure and blockwise missing values.



On this paper we focused on survival endpoints, primarily Progression-Free Survival (PFS) then Overall survival (OS) .

We developed a multimodal approach relying on a stacking algorithm as an aggregating learner from multiple sources \cite{wolpert_stacked_1992, van_der_laan_super_2007}. While late-fusion strategies have been explored for classification \cite{sklearn_api}, their extension to right-censored survival data remains under-researched. In the scikit-survival library \cite{polsterl_scikit-survival_2020}, the stacking procedure merges model outputs without applying cross-validation during aggregation, which exacerbates overfitting and leads to poor performances in external sets. Multimodality Stacking with Blockwise
missing values(MSB) decouples modality-specific learning from global aggregation, training independent base-learners on available data sources before merging their outputs via a cross-validated stacking meta-learner. This architecture provides a dual advantage: it mitigates the ``curse of dimensionality" by reducing the input space of the final aggregator and handles blockwise gaps by aggregating risk scores rather than raw features.

\section{Methods}

 \subsection{Motivation: the PIONeeR RHU biomarkers dataset}


The PIONeeR RHU (University Hospital Research) project is a nationwide multi-center prospective that screened patients suffering from advanced NSCLC treated with inhibitors of the programmed-death 1 (PD1) - programmed-death ligand 1 (PDL1) axis (\href{https://clinicaltrials.gov/study/NCT03493581}{NCT03493581}). The study was approved by the French ethics committee (Comité de Protection des Personnes Ouest II Angers, no. 2018/08) and the French drug and device regulation agency (Agence Nationale de Sécurité du Médicament, no. 2018020500208). Informed consent was obtained from each participant before any study procedure. The samples originated from biopsies, blood samples, and clinical examinations, at multiple times prior- and on-treatment.

We focus here on the pre-treatment measurements, comprising a total of 378 biomarkers. The aim of the study was to provide a better understanding of resistance to anti-PD1 therapy. The project involved 17 hospitals across the country. After exclusion criteria, there remained 437 patients available for analysis.

The data collection was organized in two steps (summarized in Figure \ref{fig:partner-role}). The samples were first collected at each clinical center and then distributed to several platforms for specific analyses, creating the multiple (8) sources of this study. The first source consisted of clinical data (source A). These are routinely recorded variables at the time patients were registered and include age, gender, and initial characteristics of the patients’ disease (e.g., histology). The second source, medical biology (source B), was also compiled independently at each medical center. It reflects the patient’s biological status more directly, as it is based on biomarkers from, e.g., blood counts or biochemistry. This information is routinely collected during patient care. More advanced circulating biomarkers originated from blood samples processed at two specialized sites: Immuno-monitoring at the immuno-profiling platform Marseille Immunopôle (or MIIPP, source C), located at the Marseille University Hospital (APHM), and vascular monitoring at the vasculomonitoring platform (or MIVBL, source D), also at APHM.

Tumor tissue samples were analyzed across three laboratories. The pathology department of APHM performed standard ImmunoHistoChemistry(IHC) analyses, including, e.g., programmed-death ligand 1 (PDL1) staining (source E). These data were divided into two sources: one for PDL1 expression  and another for mutations (source F) detected in tumor tissue (complemented by whole-exome sequencing performed by Veracyte). In addition, multiplex IHC staining was performed by Veracyte (formerly HalioDx, source G) to quantify infiltration of several (sub-)populations of immune cells, and multiplex immunofluorescence staining was carried out by InnatePharma (source H). All these sources, along with the rationale for this broad and in-depth profiling, are described in details in Barlesi et al. \cite{barlesi_integrative_2026}.

Regarding the outcome, we focused here on Progression Free survival (PFS), defined by either death of the patient or progression of the disease according to the RECIST criterion (Response Evaluation Criteria in Solid Tumors), a standard evaluation criterion for solid tumors \cite{eisenhauer_new_2009}. 
We also produced results on overall survival as an outcome in appendix.

\subsection{Multimodality Stacking with Blockwise missing values (MSB)}

The primary contribution of this work is the MSB algorithm, a meta-learning framework designed to address three concurrent challenges in clinical survival analysis: high-dimensionality, missing data and multicollinearity. It follows a late fusion paradigm, with an aggregation step explicitly accounting for the missingness pattern. It is implemented in a stacking framework . To handle these challenges, we propose to leverage the observed blockwise missingness patterns of the data.

\begin{figure}[H]
    \centering
    \includegraphics[width=1\linewidth]{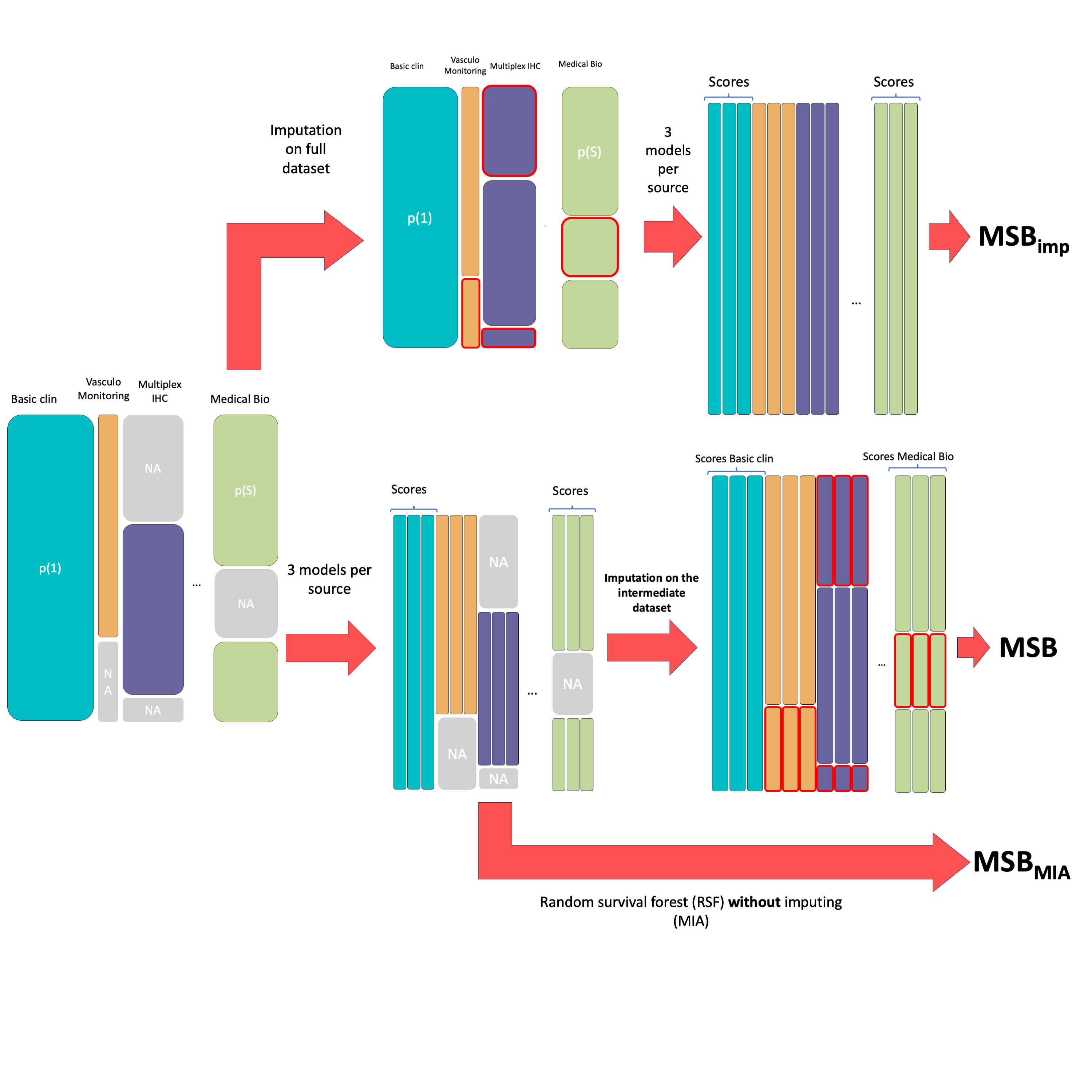}
    \caption{Multimodality Stacking with Blockwise missing values}
    \label{fig:MSB_complete}
\end{figure}

To do so, we propose three strategies. The first, straightforward, consists of imputing the original data before splitting it into modalities (MSB$_{imp}$, Figure \ref{fig:MSB_complete}). The second, analogous to the approach of van Loon et al. \cite{van_loon_imputation_2024}, performs imputation on source-aggregated survival risk scores (e.g., hazards, see below), before a stacking step (MSB).
For MSB$_{imp}$ we used a kNN (k Nearest-Neighbour) imputer with default parameters, to limit computation time while preserving predictive stability \cite{10.1093/bioinformatics/17.6.520}. For MSB, we imputed by mean.  The third strategy applies missing incorporated in attributes (MIA) directly within the meta-learner instead of using imputation (MSB$_{MIA}$, \cite{TWALA2008950}). In this approach, the missingness itself is treated as an additional category and is included in the splitting rules of the tree-based model, without replacing the missing values (Figure \ref{fig:MIAschema}). This choice was guided by the review of Josse et al. \cite{josse_consistency_2024}, which recommends MIA as suitable regardless of the underlying missingness mechanism. In our setting, this method was used when the meta-learner was a Random Survival Forest.

Our proposed approach aggregates predictions of two models within each modality: a random survival forest model, and a component-wise boosting survival model \cite{cwxgb}. The base learners are first fitted on all patients of the training dataset on the modality's features. The ``meta-learner" is then trained using the out-of-fold cross-validated predictions (risk scores) from these base estimators, to mitigate the risk of overfitting. This hierarchical approach reduces the input space of the final aggregator to a parsimonious set of modality-specific risk scores.  
Alongside the computed risk scores we appended the per-block missingness rate, since adding the missingness mask has been shown to increase performance~\cite{10.1093/gigascience/giac013}, and to check whether the algorithm learns from the missingness pattern or from biomarker values.

To ensure complete input for base learners, we applied source-specific kNN imputation: if a source contained residual missingness after accounting for blockwise patterns, a kNN imputer (with default hyperparameters) was fitted and applied independently within that source. This two-stage approach distinguishes between structural blockwise missingness (entire sources absent for patient subsets, handled by MSB's aggregation architecture) and residual feature-level gaps (handled by local imputation).

The MSB algorithm is formalized as as follows: The algorithm is split in two: the first specifies training (Algorithm \ref{alg:cap}), the second predictions (Algorithm \ref{alg:predict}).

\subsubsection{MSB pseudo code}

The complete features matrix is denoted by $X$. We denote by $n(s)$ and $p(s)$ the number of rows and columns, respectively, for the data source $s \in \left\lbrace 1,\cdots, S\right\rbrace$.  NA denotes the missing value indicator. $[X_s | X_r ]$ is the concatenation operation, an outer join on the indexes, meaning that indices present in $X_s$ but not in $X_r$ are filled with missing values in source r, and respectively for $X_r$ 

\begin{algorithm}[H]
\caption{Multimodality Stacking with Blockwise missing values algorithm: Training}\label{alg:cap}
\begin{algorithmic}
\Require 

$S$ sources of data,

$n$ total number of rows,

$X_{s} \in \mathbb{R}^{n(s) \times p(s)}$ for $s \in$ [1,S],

$y \in \mathbb{R}^{n \times 2}$, \Comment time-to-event variable and censoring indicator

$X = [X_1 | \dots | X_S] \in \big(\mathbb{R}\cup$\{NA\}\big)$^{n \times \sum_{s=1}^S p(s)}$ 

$\mathcal{M}_1,\cdots, \mathcal{M}_M$ models 

$\mathcal{ML}$ the meta-learner

Type $\in$ \{`imp', `MIA', ` '\} 

Missingness  Indicator $\in$ \{True, False\}

$R_{1}\dots R_{S}$ per-block missingness rates

\State $\hat{Z} \gets$ empty, $n$ rows
\If{Type $=$ `imp'}
    \State $X \gets$ fit and apply imputer on $X$
\EndIf

\For{s in 1:S}
        \If{if intra-source missingness in $X_s$}
        \State fit source-specific kNN imputer; impute $X_s$
        \EndIf
        \State $cv \gets$ cross validation on $X_s$
        \For{$m$ in $1,\cdots, M$}
            \State $\widehat{\mathcal{M}}_{m,s,cv} \gets$ Fit model $\mathcal{M}_m$ on in-folds of $X_s$'s cv
            \State $\hat{Z}_{m, s} \gets$ out of fold predictions of $\widehat{\mathcal{M}}_{m,s,cv}$  concatenated by row 
        \EndFor
        \State $\hat{Z} \gets[\hat{Z} | \hat{Z}_{1, s}|\dots|\hat{Z}_{M, s}]$ 
        \For{$m$ in $1,\cdots, M$}
            \State $\widehat{\mathcal{M}}_{m,s} \gets$ Fit model $\mathcal{M}_m$ on $X_s$ 
        \EndFor
\EndFor
\If{Missingness Indicator}
    \State $\hat{Z} \gets [\hat{Z} \mid R_{1} \mid \dots \mid R_{S}]$
\EndIf
\If{Type $\neq$ `MIA'}
    \State $\hat{Z} \gets$ fit and apply imputer on $\hat{Z}$
\EndIf

\State $\widehat{\mathcal{ML}} \gets$ fit $\mathcal{ML}$ on $\hat{Z}$
\State \Return $\widehat{\mathcal{ML}},\ \{\widehat{\mathcal{M}}_{m,s}, m \in [1,M],s\in[1,S]\}$

\end{algorithmic}
\end{algorithm}

For a new set of features $X_{new}$, the prediction is the one of $\widehat{\mathcal{ML}}$ on the predictions of each $M \times S$ fitted models (each $\widehat{\mathcal{M}}_{m,s}$) on $X_{new}$.

\begin{algorithm}
\caption{Multimodality Stacking with Blockwise missing values algorithm: Predictions}\label{alg:predict}
\begin{algorithmic}
\Require 

$S$ sources of data,

$n$ total number of rows,

$X_{s} \in \mathbb{R}^{n(s) \times p(s)}$ for $s \in$ [1,S],

$X = [X_1| \dots |X_S] \in \big(\mathbb{R}\cup$\{NA\}\big)$^{n \times \sum_{s=1}^S p(s)}$ 

$\{\widehat{\mathcal{M}_{m,s}}, m \in [1,M],s\in[1,S]\}$ Fitted models,

$\widehat{\mathcal{ML}}$ a fitted meta-learner

Missingess Indicator : True or False 

$R_{1} \dots R_{S}$ The rate of missingness per block

\State $\hat{Z} \gets$ empty, $n$ rows
\If{Type $=$ `imp'}
    \State $X \gets$ apply fitted imputer on $X$
\EndIf

\For{s in 1:S}
        \If{if intra-source missingness exists in $X_s$ }
        \State Impute Xs's missing values on saved source specific kNN imputer
        \EndIf
        \For{m in 1:M}
        \State $Z_{m,s} \gets \widehat{\mathcal{M}_{m,s}}(X_s)$
        \EndFor
        \State $\hat{Z} \gets [\hat{Z} | \hat{Z}_{1, s}|\dots|\hat{Z}_{M, s}] $

\EndFor

\If{Missingness Indicator}
    \State $\hat{Z} \gets [\hat{Z} \mid R_{1} \mid \dots \mid R_{S}]$
\EndIf
\If{Type $\neq$ `MIA'}
    \State $\hat{Z} \gets$ apply fitted imputer on $\hat{Z}$
\EndIf

\Return  $\widehat{\mathcal{ML}}(\hat{Z})$
\end{algorithmic}
\end{algorithm}


\subsection{Metrics}

The performance of each model was assessed using five repetitions of a stratified 5-fold cross-validation (\texttt{RepeatedStratifiedKFold}), stratified on the event indicator, giving 25 train/test folds. This provides a more reliable performance estimate while keeping computational cost reasonable. All preprocessing (imputation, block-wise fitting) was performed on the training fold and applied to the held-out fold. For each fold, training used Algorithm~\ref{alg:cap} and prediction Algorithm~\ref{alg:predict}. We used scikit-survival~\cite{polsterl_scikit-survival_2020} with the scikit-learn API~\cite{sklearn_api}.

To assess whether MSB's improvements were systematic rather than attributable to random variation across cross-validation folds, we conducted paired Wilcoxon signed-rank tests comparing each MSB variant to its corresponding baseline. Wilcoxon tests were chosen over parametric $t$-tests due to the bounded nature of the C-index. $p$-values were adjusted for multiple comparisons using the Holm procedure.

\subsection{Benchmarking}\label{sec:benchmarking}

The SurvBoard benchmark \cite{wissel_survboard_2025} provides a standardized framework for comparing multi-omics survival models, which we used to situate MSB relative to the current state of the art. It comprises 21 multi-omics survival cohorts from The Cancer Genome Atlas (TCGA), with fixed cross-validation splits and publicly released prediction files. These cohorts are then used on 12 models on those cross validation folds. By fixing the data splits and preprocessing, the benchmark removes analyst-dependent choices that might otherwise artificially boost reported performance. We ran MSB on SurvBoard’s predefined folds and evaluated its predictions using the benchmark’s own metric implementations — time-dependent (Antolini) concordance \cite{antolini_time-dependent_2005} and integrated Brier score on the time grid specified in the repository — making MSB’s outcomes directly comparable to the published leaderboard rather than re-estimated under alternative conventions.

The 21 TCGA cohorts are referenced on the study abbreviations \cite{tomczak_cancer_2015}  (e.g., BRCA for breast invasive carcinoma; KIRC for kidney renal clear cell carcinoma).

Each cohort includes up to seven modalities — clinical variables, gene expression (GEX), copy number variation (CNV), DNA methylation (meth), microRNA expression (miRNA), mutation status (mut), and reverse-phase protein array (RPPA) — with seven of the 21 cohorts lacking one or more.

For omics blocks whose feature dimensionality exceeds that of other sources by roughly an order of magnitude (notably GEX and meth), we applied PCA and retained the top 20 components per source before training the base learners, fitting the PCA only within the training folds. This projects each very high-dimensional block into a compact, low-rank representation, limiting the noise carried forward into the source-level risk estimates. Compressing high-dimensional omics data is an active research area, ranging from linear projections such as PCA to pathway-informed transformers \cite{liu_pathformer_2024}; we chose PCA for its simplicity. A sweep over $k \in \{5, 20, 40\}$ showed $k=5$ underperformed while $k=40$ gave no further gain, confirming 20 components as a stable operating point. This within-source reduction was independent of MSB's fusion-stage compression, which restricts the aggregator's inputs to a small set of per-source risk scores regardless of block dimensionality.

SurvBoard additionally provides training sets with missing modalities while keeping test sets complete, enabling fair cross-model comparison. We trained MSB on these partially observed sets as well, since handling blockwise missingness is a primary design objective of our method.

\subsubsection*{Performance}
To compare the algorithms, we used the C-index and the integrated Brier Score (iBS) (\cite{harrell_multivariable_1996,graf_assessment_1999}). The purpose was to compare algorithms both at ranking risk (C-index) and at predicting true survival (iBS). When examining the iBS scores, we aimed to determine whether MSB algorithms performed better in the short term (to see which algorithm could differentiate best the early progressors from the others) or/and in the long term. Therefore, two iBS were calculated: 1) from day 10 to day 100 and 2) from day 100 until day 1000 in the results section, namely `iBS Early' and `iBS Late'.


\subsubsection*{Modality importance}


We also investigated the behavior of each algorithm at the modality level. To do so, we used permutation importance \cite{perm_importance} as a way to measure the modality importance. The metric used for permutation importance was based on the iBS but for readability we adjusted it so that `the higher the better', ideally ranging from 0 to 1 (an integrated Brier score can be worse than random and therefore being above 0.25). We thus defined:

$$
iBSS = 1 - \frac{iBS}{0.25},
$$

a Brier Skill Score, representing the level of improvement of a Brier score compared to a reference (here 0.25, the iBS obtained from random decision). 

Using this metric, we computed permutation importance per source rather than per (source, model) pair, since the risk scores produced by different base learners on the same source are highly correlated. This grouped formulation follows Gregorutti et al.~\cite{gregorutti_grouped_2015}, who show that jointly permuting correlated variables is necessary to avoid the underestimation of importance that affects individual permutation under correlation. We did not adopt their random-forest-based group-selection procedure, as our groups are fixed a priori by the source structure. Formally, for each source $s$:
$$
perm_s \;=\; iBSS\left(\hat{f}(\mathbf{Z})\right)
- \frac{1}{K}\sum_{k=1}^{K}
iBSS\left(\hat{f}\!\left(\mathbf{Z}\,;\,\tilde{\mathbf{z}}^{(s)}_{\pi_k}\right)\right)
$$

where $\hat{f}$ is the fitted meta-learner, $\mathbf{Z}$ is the full risk-score
matrix, $\tilde{\mathbf{z}}^{(s)}_{\pi_k}$ denotes the block of source $s$'s
scores under the $k$-th joint row-permutation $\pi_k$, and $K$ is the number of
permutations.

We measured modality importance by jointly permuting the 2 risk scores of each of the 8 sources, with $K=100$ permutations

\section{Results}

\subsection{Benchmarking MSB against other multisource models}

\begin{figure}[H]
    \centering
    \begin{subfigure}[t]{0.49\linewidth}
        \centering
        \includegraphics[width=\linewidth]{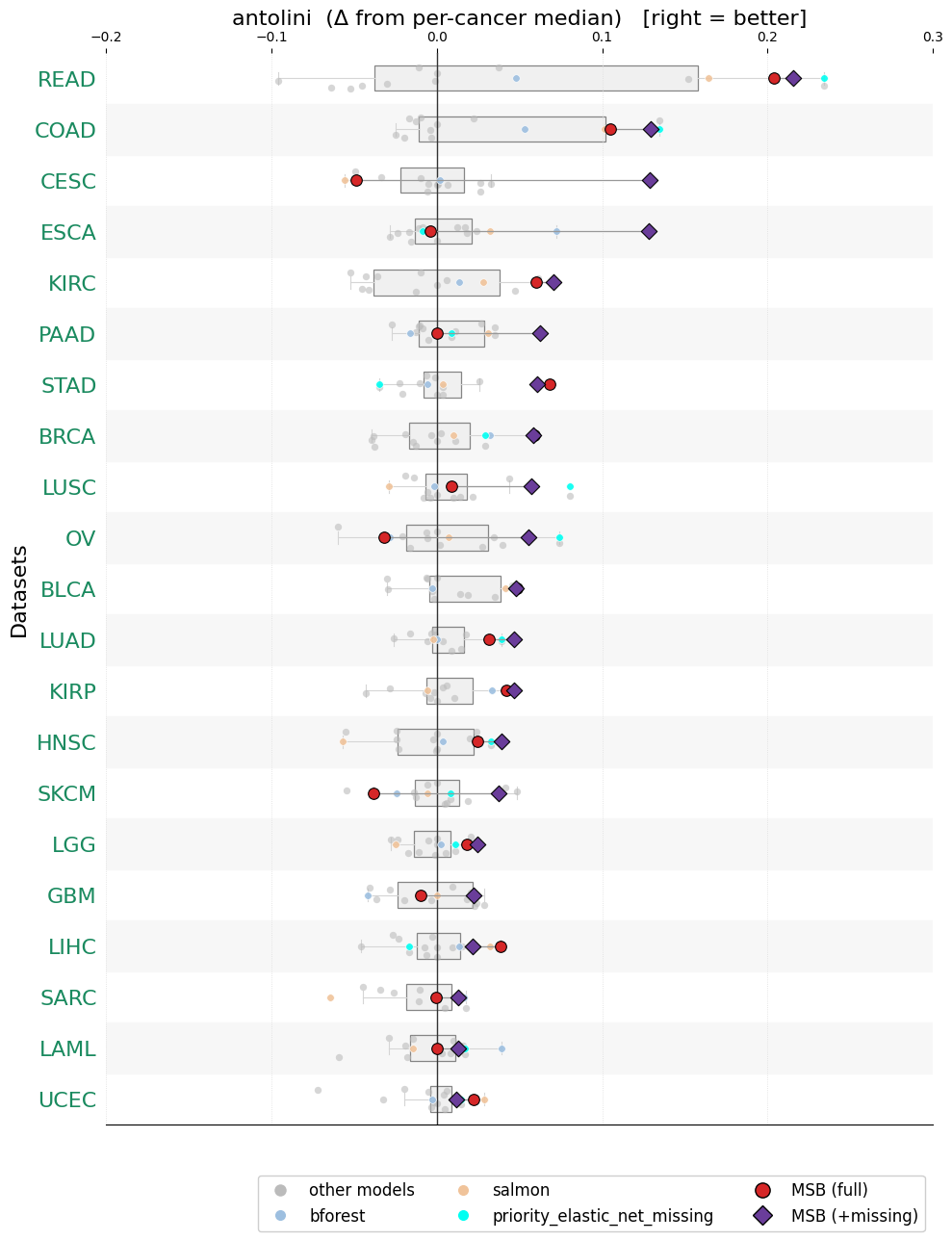}
        \caption{Antolini's c-index (higher is better).}
        \label{fig:cindex}
    \end{subfigure}
    \hfill
    \begin{subfigure}[t]{0.49\linewidth}
        \centering
        \includegraphics[width=\linewidth]{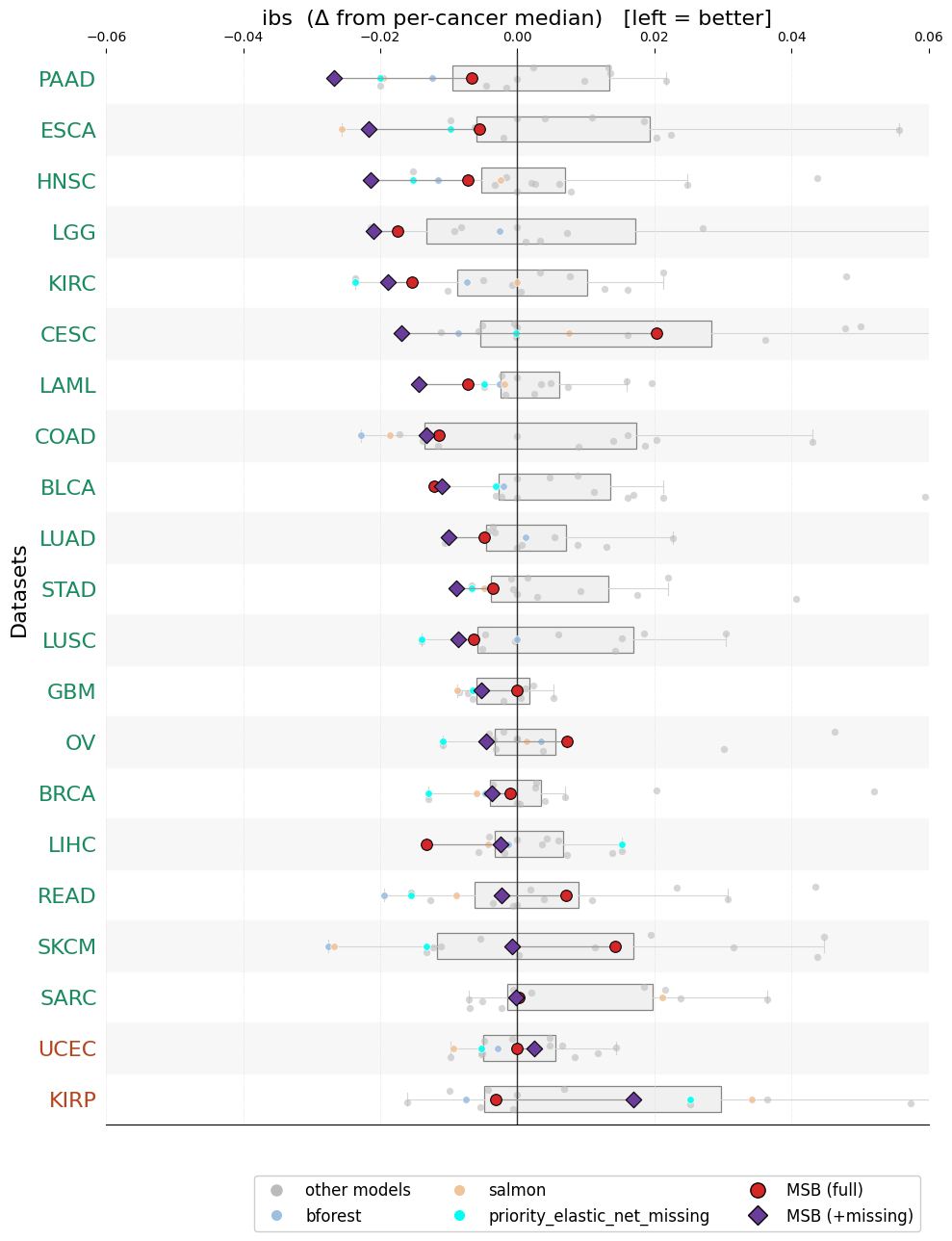}
        \caption{Integrated Brier Score (lower is better).}
        \label{fig:ibs}
    \end{subfigure}
    \caption{MSB (complete-modality and with-missing-values variants) benchmarked against the 13 SurvBoard multi-omics models across 21 TCGA cohorts. Boxplots show each cohort's field distribution; points give each model's mean value (centered around the median) across the 25 official folds. Dataset names are green where MSB (with missing) exceeds the field median. (a) Antolini's c-index, cohorts ordered by descending field median (leftmost: LGG, KIRP, UCEC, KIRC, most predictable; rightmost: LUSC, SKCM, PAAD, near-random, median $\sim$ 0.5). (b) Integrated Brier Score, cohorts ordered by descending field median (hardest-to-calibrate leftmost).}
    \label{fig:benchmark}
\end{figure}

The benchmark's results on SurvBoard datasets are shown in Figures~\ref{fig:cindex} and~\ref{fig:ibs}, comparing both MSB (complete-modality) and MSB (with missing values) against the full field, with BlockForest \cite{hornung_block_2019}, SALMON \cite{huang_salmon_2019} and priority elastic net (with missing values) \cite{musib_priority-elastic_2024} highlighted for reference.

On discrimination, MSB (complete-modality) ranked first by median Antolini concordance across the 21 cohorts, sitting at or near the best-performing method in the majority of them (e.g. KIRP, UCEC, KIRC, SARC, BRCA, LUAD, LIHC, HNSC, READ). BlockForest and SALMON tracked MSB closely on high-signal cohorts but fell behind on several mid-table ones. On calibration, MSB (complete-modality) ranked third by median IBS, within a negligible margin of the two leading methods (difference$\approx 0.0035$). Substituting the stacking meta-learner (RSF, CoxNet, ComponentWise Boosting) caused minor rank reshuffling but left aggregate performance essentially unchanged (Appendix), indicating the result is not an artifact of a single meta-learner choice.

BlockForest is the closest architectural counterpart to MSB — an intermediate-fusion analogue to MSB's late fusion — while requiring none of the sample size demanded by neural models. SALMON is better understood as a learned dimensionality-reduction front-end atop a Cox objective rather than a distinct fusion architecture; that MSB, using only unsupervised PCA, matches SALMON's discrimination suggests the survival-relevant within-source signal is largely low-rank on these cohorts. This complete-modality benchmark establishes that MSB is competitive with state-of-the-art multi-omics survival models under standard conditions, and justifies applying it to the setting it was designed for: blockwise missingness.

Training MSB on SurvBoard's partially observed datasets increased the number of training observations per cohort, in some cases substantially (from 80 up to 451), while the 25-fold test sets remained unchanged. In this setting, MSB ranked first on both discrimination and calibration across the field (Table~\ref{tab:survboard_rank}), confirmed by the Friedman test — an improvement over the third-place calibration rank of the complete-modality variant (Table \ref{tab:survboard_rank}). This is the setting the method targets: rather than treating blockwise missingness as a nuisance to be imputed away or a subset of patients to be discarded, MSB uses the additional partially observed patients directly during training, and the gain over the complete-modality variant indicates this additional information is being exploited rather than merely tolerated.

This initial step demonstrates that the algorithm is appropriate and that applying it to our target cohort, the PIONeeR dataset, is justified.

\subsection{Results on the PIONeeR dataset}
\subsubsection{Missing value patterns in PIONeeR}

\begin{figure}[H]
    \centering
    \includegraphics[width=1\linewidth]{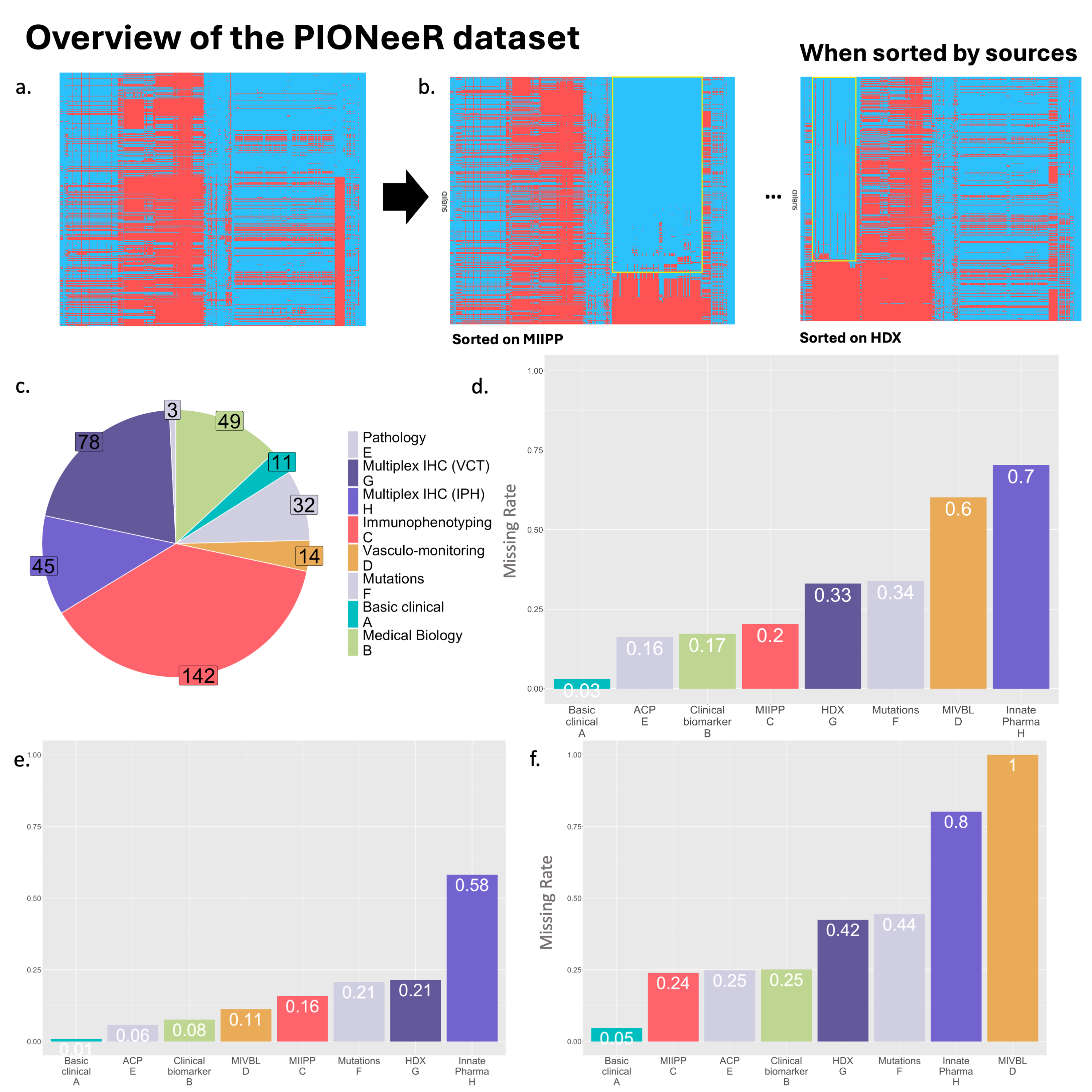}
    \caption{PIONeeR's data overview, multiple sources at baseline :\\ To visualize these patterns, we sorted the data using a binary indicator based on a 50\% missingness threshold per source. This arrangement highlights the blockwise missingness structure, where specific patient cohorts are associated with the presence or absence of entire data modalities.
    a: General overview of missingness in the dataset,b: Example of blockwise missingness in the PioNeer dataset when sorted by sources, the data is blue when available and red when missing, c: Number of features per sources, d: Rates of missing values' sources on full data, e: Rates of missing values' sources on APHM patients , f: Rates of missing values' sources from other centers}
    \label{fig:sources_Baseline}
\end{figure}

The PIONeeR dataset comprises 378 variables for 437 observed patients. The collaborative work of the project partners (Figure \ref{fig:partner-role}), while providing information allowing for a deep exploration to explain the progression of the disease, also generated missing data (39\%, Figure \ref{fig:sources_Baseline}a). The latter could be explained by two main causes. The first is logistical: some samples were not planned to be extracted for some patients (e.g., patients from a specific clinical center). The second was due to the patients' sample availability when the analysis was conducted.

Sorting the aggregated data by source revealed that the missingness mechanism was mostly at the modality level (`blockwise missingness', Figure \ref{fig:sources_Baseline}b). There was a distinct missingness mechanism for each source but at the source level, each feature seemed to share the same or almost the same missingness mechanism as other features from the same source. However, some variables also exhibited intra-block missingness—individual gaps within an otherwise available modality.

The number of features and missingness rates varied significantly across sources respectively from 3 to 142 features and from 3\% (basic clinical) to 70\% missing rate (Figure \ref{fig:sources_Baseline}c\ \& d). Basic clinical data (Source A) comprised demographics, tobacco history, and tumor characteristics (e.g., histology and number of metastases) collected during the screening visit. These features were nearly complete ($<5\%$ missingness), reflecting their role in routine clinical care. A critical variable in this modality was the treatment line: first-line patients received a combination of chemotherapy and ICI, whereas subsequent-line patients were treated with ICI monotherapy. As Source A integrates a broad range of general characteristics, it represents the most heterogeneous – yet most readily available – data modality in the cohort.

Among the blood-derived modalities (Sources B, C, and D), missingness rates varied significantly due to logistical stratification. While clinical biochemistry and inflammatory markers (Sources B and C) were largely available across the cohort, vasculo-monitoring (Source D) was exclusively implemented at the coordinating center (APHM), which accounted for 45\% of the total cohort (Figure \ref{fig:sources_Baseline}e). Consequently, this source was structurally absent for patients enrolled across the other 16 participating centers (Figure \ref{fig:sources_Baseline}f). To assess whether this center-specific data availability introduced a prognostic bias, we compared Kaplan-Meier survival curves between APHM and non-APHM patients (Figure \ref{fig:KMPLot}). The absence of a statistically significant difference between the two curves indicated that the two patient subgroups had comparable baseline survival outcomes and that MSB's predictions were unlikely to be confounded by center identity.

To account for distinct data generation processes, we separated the data derived from pathology samples into Source E (PD-L1 expression) and Source F (mutations). This was justified by their disparate missingness rates (16\% vs. 34\%), which reflected the difference between standard immunohistochemistry and complex genomic sequencing.

Eventually, the tumor-derived modalities exhibited the highest rates of missingness. This sparsity was inherent to the complexity of the PIONeeR protocol \cite{barlesi_integrative_2026}.
Consequently, feature availability in these sources was directly proportional to the quantity and quality of the biopsied material.

To assess whether blockwise missingness was associated with patient prognosis, we ran log-rank tests per source. Six of eight showed no significant association ($p>0.05$); the two nominal associations (Sources C and F) were no significant when stratified by treatment line (Fig.~\ref{fig:S4}).

\subsubsection{Enhanced PFS Prediction via MSB}

We configured MSB to process each of the eight data sources independently. As a result, the meta-learner was trained on a consolidated matrix of 24 input features ($8 \text{ modalities} \times (2 \text{ base models}+ $block missingness rate$)$). This reduced feature space was intentionally chosen to ensure the model remained parsimonious, providing a stable ratio of observations to predictors for the aggregating learner. To assess the effectiveness of the MSB framework, we compared its performance with baseline survival models trained on the kNN-imputed feature set (Figure \ref{fig:MSB_complete}). 

\begin{figure}[H]
    \centering
    \includegraphics[width=1\linewidth]{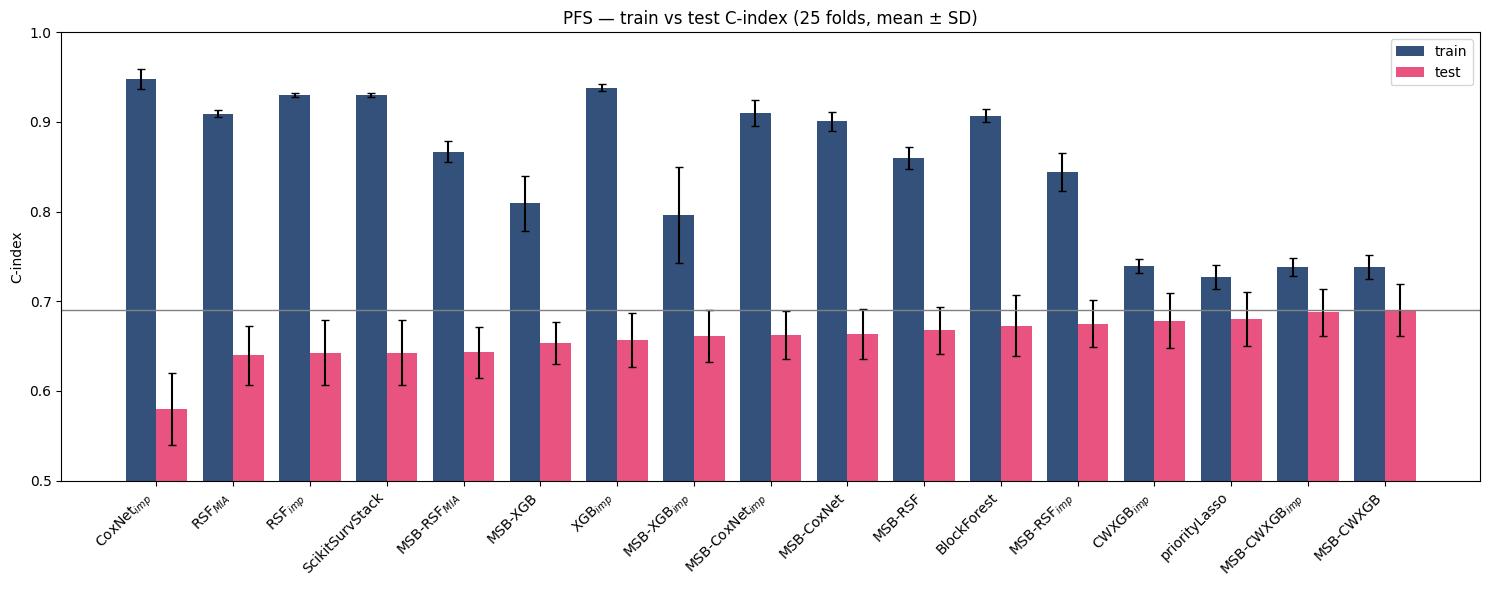}
    \caption{Summary of the results from \ref{tab:performance} for C-index obtained sorted by their test values. This figure summarizes the table  how well MSB reduces overfitting and improves performances. The line displayed is the highest mean C-index achieved on test on PFS as outcome.}
    \label{fig:cindexsummary}
\end{figure}

Across architectures — CoxNet, Component-Wise Gradient Boosting (CWXGB) and Random Survival Forest (RSF), though not Gradient Boosting (XGB) — MSB showed higher discriminative power than the corresponding baseline survival algorithms (Tables~\ref{tab:performance} and~\ref{tab:msb_validation_cindex}). Figure\ref{fig:cindexsummary} displays a graphical representation of the results.

 MSB was also evaluated on this imputed dataset (denoted by the label `imp'). The `imp' version of MSB maintained its hierarchical structure, learning modality-specific representations separately prior to meta-aggregation. The performance of MSB remained consistent whether trained on the original sparse modalities or the imputed data. This suggests that, for this cohort, the primary challenge lies in managing high dimensionality rather than missingness alone.

By breaking down the overall feature space into separate, modality-focused sub-tasks, MSB successfully reduced the complexity arising from the PIONeeR dataset’s heterogeneous sources. This reduction in complexity did not degrade performance; in fact, it led to improved results.

We also compared MSB to selected models from the SurvBoard benchmark to position it among other block-aware algorithms. We chose BlockForest and priorityLasso, which showed strong performance in that benchmark.

These results indicate that for multi-source clinical data characterized by structural sparsity and high dimensionality, a structured late-fusion strategy is well suited to survival estimation. They also point to the importance of the meta-learner, since performance varied substantially across MSB configurations.

\begin{table}[H]
\centering
\caption{Performance table for the PFS outcome (5-fold CV, 5 repetitions). Bold indicates family best; $^*$ indicates global best (Test only). All models include missing block indicators. CWXGB = Component-Wise Gradient Boosting, RSF = Random Survival Forest, XGB = Gradient Boosting, MSB = Multimodality Stacking with Blockwise missing values model.}
\label{tab:performance}
\begin{tabular}{lllrrr}
\toprule
Model Family & Model & Dataset & C-index ($\pm$ SD) & iBS Early & iBS Late \\
\midrule
Linear & MSB-CoxNet$_{imp}$ & test & 0.663\,{\scriptsize$\pm$\,0.027} & 0.112\,{\scriptsize$\pm$\,0.010} & 0.225\,{\scriptsize$\pm$\,0.019} \\
 &  & {\scriptsize train} & {\scriptsize 0.910\,$\pm$\,0.014} & {\scriptsize 0.077\,$\pm$\,0.009} & {\scriptsize 0.088\,$\pm$\,0.012} \\
 & MSB-CoxNet & test & 0.664\,{\scriptsize$\pm$\,0.028} & 0.112\,{\scriptsize$\pm$\,0.010} & 0.227\,{\scriptsize$\pm$\,0.017} \\
 &  & {\scriptsize train} & {\scriptsize 0.901\,$\pm$\,0.010} & {\scriptsize 0.080\,$\pm$\,0.007} & {\scriptsize 0.088\,$\pm$\,0.009} \\
 & CoxNet$_{imp}$ & test & 0.580\,{\scriptsize$\pm$\,0.040} & 0.214\,{\scriptsize$\pm$\,0.034} & 0.350\,{\scriptsize$\pm$\,0.042} \\
 &  & {\scriptsize train} & {\scriptsize 0.948\,$\pm$\,0.011} & {\scriptsize 0.025\,$\pm$\,0.004} & {\scriptsize 0.041\,$\pm$\,0.007} \\
 & priorityLasso & test & \textbf{0.680\,{\scriptsize$\pm$\,0.030}} & \textbf{0.101\,{\scriptsize$\pm$\,0.013}} & \textbf{0.196\,{\scriptsize$\pm$\,0.016}} \\
 &  & {\scriptsize train} & {\scriptsize 0.727\,$\pm$\,0.014} & {\scriptsize 0.091\,$\pm$\,0.004} & {\scriptsize 0.169\,$\pm$\,0.007} \\
\midrule
Forest & MSB-RSF & test & 0.668\,{\scriptsize$\pm$\,0.026} & \textbf{0.100\,{\scriptsize$\pm$\,0.010}$^*$} & \textbf{0.204\,{\scriptsize$\pm$\,0.010}} \\
 &  & {\scriptsize train} & {\scriptsize 0.859\,$\pm$\,0.012} & {\scriptsize 0.065\,$\pm$\,0.002} & {\scriptsize 0.123\,$\pm$\,0.008} \\
 & MSB-RSF$_{imp}$ & test & \textbf{0.675\,{\scriptsize$\pm$\,0.026}} & 0.102\,{\scriptsize$\pm$\,0.009} & 0.205\,{\scriptsize$\pm$\,0.012} \\
 &  & {\scriptsize train} & {\scriptsize 0.844\,$\pm$\,0.021} & {\scriptsize 0.070\,$\pm$\,0.003} & {\scriptsize 0.143\,$\pm$\,0.013} \\
 & MSB-RSF$_{MIA}$ & test & 0.643\,{\scriptsize$\pm$\,0.028} & 0.108\,{\scriptsize$\pm$\,0.010} & 0.218\,{\scriptsize$\pm$\,0.010} \\
 &  & {\scriptsize train} & {\scriptsize 0.867\,$\pm$\,0.012} & {\scriptsize 0.074\,$\pm$\,0.003} & {\scriptsize 0.125\,$\pm$\,0.010} \\
 & RSF$_{imp}$ & test & 0.643\,{\scriptsize$\pm$\,0.036} & 0.105\,{\scriptsize$\pm$\,0.011} & 0.210\,{\scriptsize$\pm$\,0.012} \\
 &  & {\scriptsize train} & {\scriptsize 0.930\,$\pm$\,0.003} & {\scriptsize 0.041\,$\pm$\,0.001} & {\scriptsize 0.054\,$\pm$\,0.001} \\
 & RSF$_{MIA}$ & test & 0.640\,{\scriptsize$\pm$\,0.033} & 0.108\,{\scriptsize$\pm$\,0.013} & 0.209\,{\scriptsize$\pm$\,0.007} \\
 &  & {\scriptsize train} & {\scriptsize 0.909\,$\pm$\,0.004} & {\scriptsize 0.055\,$\pm$\,0.002} & {\scriptsize 0.087\,$\pm$\,0.003} \\
 & BlockForest & test & 0.673\,{\scriptsize$\pm$\,0.034} & 0.103\,{\scriptsize$\pm$\,0.012} & 0.207\,{\scriptsize$\pm$\,0.007} \\
 &  & {\scriptsize train} & {\scriptsize 0.907\,$\pm$\,0.007} & {\scriptsize 0.050\,$\pm$\,0.002} & {\scriptsize 0.101\,$\pm$\,0.002} \\
\midrule
Boosting & MSB-CWXGB & test & \textbf{0.690\,{\scriptsize$\pm$\,0.029}$^*$} & 0.102\,{\scriptsize$\pm$\,0.014} & \textbf{0.195\,{\scriptsize$\pm$\,0.010}$^*$} \\
 &  & {\scriptsize train} & {\scriptsize 0.738\,$\pm$\,0.014} & {\scriptsize 0.093\,$\pm$\,0.004} & {\scriptsize 0.171\,$\pm$\,0.005} \\
 & MSB-CWXGB$_{imp}$ & test & 0.688\,{\scriptsize$\pm$\,0.027} & \textbf{0.102\,{\scriptsize$\pm$\,0.014}} & 0.197\,{\scriptsize$\pm$\,0.010} \\
 &  & {\scriptsize train} & {\scriptsize 0.738\,$\pm$\,0.010} & {\scriptsize 0.093\,$\pm$\,0.003} & {\scriptsize 0.171\,$\pm$\,0.005} \\
 & CWXGB$_{imp}$ & test & 0.679\,{\scriptsize$\pm$\,0.031} & 0.105\,{\scriptsize$\pm$\,0.014} & 0.201\,{\scriptsize$\pm$\,0.010} \\
 &  & {\scriptsize train} & {\scriptsize 0.739\,$\pm$\,0.008} & {\scriptsize 0.097\,$\pm$\,0.002} & {\scriptsize 0.175\,$\pm$\,0.003} \\
 & MSB-XGB & test & 0.654\,{\scriptsize$\pm$\,0.023} & 0.106\,{\scriptsize$\pm$\,0.010} & 0.223\,{\scriptsize$\pm$\,0.015} \\
 &  & {\scriptsize train} & {\scriptsize 0.810\,$\pm$\,0.031} & {\scriptsize 0.076\,$\pm$\,0.008} & {\scriptsize 0.147\,$\pm$\,0.019} \\
 & MSB-XGB$_{imp}$ & test & 0.661\,{\scriptsize$\pm$\,0.029} & 0.106\,{\scriptsize$\pm$\,0.014} & 0.221\,{\scriptsize$\pm$\,0.021} \\
 &  & {\scriptsize train} & {\scriptsize 0.797\,$\pm$\,0.054} & {\scriptsize 0.081\,$\pm$\,0.010} & {\scriptsize 0.156\,$\pm$\,0.029} \\
 & XGB$_{imp}$ & test & 0.657\,{\scriptsize$\pm$\,0.030} & 0.105\,{\scriptsize$\pm$\,0.014} & 0.228\,{\scriptsize$\pm$\,0.017} \\
 &  & {\scriptsize train} & {\scriptsize 0.938\,$\pm$\,0.004} & {\scriptsize 0.048\,$\pm$\,0.002} & {\scriptsize 0.068\,$\pm$\,0.003} \\
\bottomrule
\end{tabular}
\end{table}

\begin{table}[H]
\centering
\caption{MSB framework validation on discrimination: paired Wilcoxon signed-rank tests comparing each MSB variant to its baseline across the 25 cross-validation folds (test sets, C-index). Adjusted $p$-values use Holm correction. \textit{Wins} is the number of folds (of 25) in which MSB achieved a higher C-index than the baseline; because repeated cross-validation folds are not independent, $p$-values are optimistic and are reported alongside win counts as consistency indicators.}
\label{tab:msb_validation_cindex}
\begin{tabular}{lcccc}
\toprule
\textbf{Comparison} & \textbf{$\Delta$ C-index} & \textbf{Wins} & \textbf{$p$} & \textbf{Adj. $p$} \\
\midrule
MSB-CoxNet vs CoxNet       & $+0.084$ ($+14.4\%$) & 23/25 & $<10^{-4}$ & $<10^{-4}$*** \\
MSB-RSF$_{imp}$ vs RSF             & $+0.032$ ($+5.0\%$)  & 24/25 & $<10^{-4}$ & $<10^{-4}$*** \\
MSB-CWXGB vs CWXGB         & $+0.012$ ($+1.8\%$)  & 18/25 & $0.013$    & $0.037$* \\
MSB-XGB vs XGB             & $-0.003$ ($-0.5\%$)  & 10/25 & $0.52$     & $1.00$ (n.s.) \\
MSB-CWXGB vs BlockForest   & $+0.018$ ($+2.6\%$)  & 20/25 & $0.0008$   & $0.004$** \\
MSB-CWXGB vs PriorityLasso & $+0.010$ ($+1.5\%$)  & 17/25 & $0.004$    & $0.015$* \\
MSB-RSF$_{imp}$ vs BlockForest     & $+0.002$ ($+0.3\%$)  & 15/25 & $0.67$     & $1.00$ (n.s.) \\
\bottomrule
\multicolumn{5}{l}{\footnotesize * $p<0.05$, ** $p<0.01$, *** $p<0.001$ (Holm-corrected). Positive $\Delta$: MSB improved (higher) C-index.} \\
\end{tabular}
\end{table}

The results of tables \ref{tab:performance} above can be summarized graphically, as shown in figure \ref{fig:cindexsummary}.

\subsubsection{Statistical validation of the MSB Framework}

We compared each MSB variant to its corresponding unified baseline using paired Wilcoxon signed-rank tests across the 25 cross-validation folds, with Holm correction (Table~\ref{tab:msb_validation_cindex}). 
For discrimination, MSB improved on the baseline in every family except gradient boosting. The gain was largest for the linear model (MSB-CoxNet 0.664 vs CoxNet 0.580, $+14.4\%$, 23 of 25 folds, $p<0.001$), moderate for random survival forests (MSB-RSF$_{imp}$ 0.675 vs RSF$_{imp}$ 0.643,
$+5.0\%$, 24 of 25 folds, $p<0.001$), and smallest for component-wise boosting
($+1.8\%$, 18 of 25 folds, $p=0.013$; $p=0.037$ after Holm correction).

 Performance was closely related to baseline strength across all metrics. (Table~\ref{tab:performance}). However source level decomposition (trough MSB) improved results.

The imputation variants MSB$_{imp}$ and MSB gave near-identical results within each family ($\Delta$C-index $<0.01$), indicating the choice between imputing raw features and imputing risk scores had little effect. The MIA variant, which forgoes imputation entirely, performed below both (e.g., MSB-RSF$_{MIA}$ 0.643 vs MSB-RSF$_{imp}$ 0.675), suggesting that, for this cohort, explicit imputation was preferable to handling missingness within the model.

\subsubsection{Comparison with BlockForest and priorityLasso}

MSB-CWXGB improved on BlockForest in discrimination (0.690 vs 0.673, $+2.6\%$, 20 of 25 folds, $p=0.0008$; $p=0.004$ after Holm correction) and long-term calibration (iBS Late 0.195 vs 0.207, 22 of 25 folds, $p<0.001$), while short-term calibration was indistinguishable.

MSB narrowed the train--test C-index gap relative to its own baseline in every family: 0.247 vs 0.368 (CoxNet), 0.169 vs 0.287 (RSF), 0.050 vs 0.060 (CWXGB), and 0.136 vs 0.281 (XGB), all for the $_{imp}$ variants under default hyperparameters. The effect is largest where the baseline overfits most, and the absolute gap tracks the base learner's own regularization ; smallest for component-wise boosting, largest for CoxNet. BlockForest's gap (0.234) sat between the two, below a standard RSF (0.287) but above MSB's forest variant; priorityLasso, with a gap of 0.047, was the least overfit model in the study.

PriorityLasso performed close to MSB's best configuration (C-index 0.680 vs 0.690) and ahead of most other models, indicating that a linear, block-aware approach is well suited to this problem. Several aspects of its fitting procedure parallel MSB: block-level coefficients are learned from out-of-fold predictions, in a stacking-like manner, and each block's contribution is shaped first by its position in the priority order and then by the penalty term. The main practical difference is that PriorityLasso requires this priority order to be specified in advance. We used the natural ordering of the sources (A--H); permuting it lowered the C-index to 0.674 and raised iBS Late from 0.196 to 0.198, indicating a modest sensitivity to a choice MSB does not have to make.

\subsubsection{Modality importance}

As described in the methods section, we computed the modality importance for MSB-CWXGB

\begin{figure}[H]
    \centering
    \includegraphics[width=0.8\linewidth]{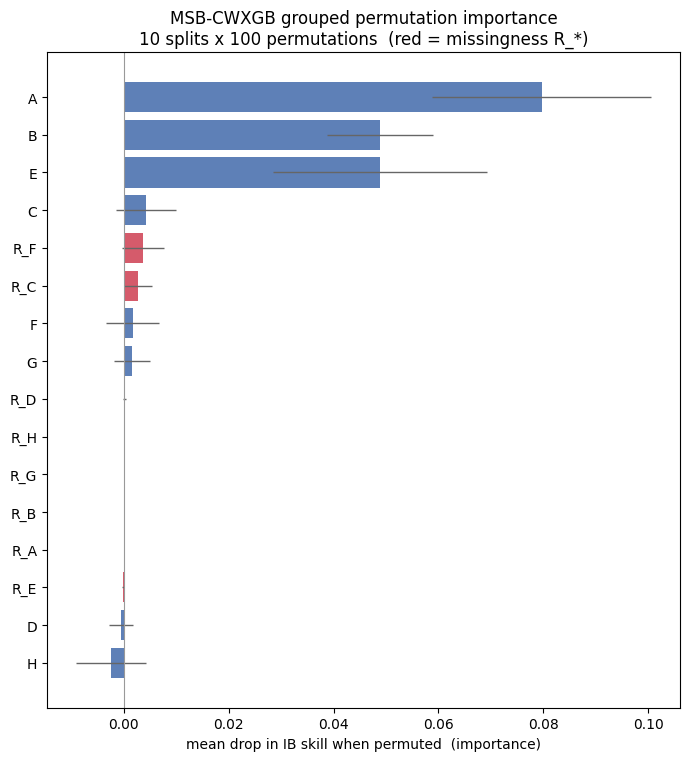}
    \caption{Top Permutation importance on iBSS for ComponentWise boosting as meta model in MSB. The y-axis corresponds to the source and x-axis the permutation importance measured as described in the Methods section}
    \label{fig:imp_PFS_cwxgb}
\end{figure}

For the MSB-CWXGB configuration—which achieved the highest predictive accuracy—three modalities emerged as the dominant predictors (Figure~\ref{fig:imp_PFS_cwxgb}): basic clinical features (Source A), medical biology (Source B), and PD-L1 Expression (Source E). These three sources consistently exhibited the highest permutation importance across 10 splits with 100 permutations. 

The dominance of Sources A, B, and E persisted across meta-learner architectures. CoxNet and RSF as meta-learners produced a similarly concentrated importance profile (Figure~\ref{fig:imp_PFS_CoxNet},~\ref{fig:imp_PFS_RSF})

In addition, the missing-block indicators showed negligible direct contribution to the model. While missingness does indirectly influence the model through imputation (e.g., neighbor selection in kNN, pooling of available risk scores, or through median imputation), the low permutation importance demonstrates that knowing which blocks are missing added minimal prognostic information beyond what the observed features already provided. 

The near-zero importance of the missing-block indicators, together with the non-significant Kaplan–Meier differences across missingness strata (Figures~\ref{fig:S4},~\ref{fig:S5}), is consistent with an absence of informative missingness bias in PIONeeR: the absence of a modality, such as tumor-based markers, does not appear to act as a proxy for prognosis.

Although there is inherent variability between data splits, the repeated selection of clinical, biological, and PD-L1 markers highlights the robustness of these features ,while tumor-based modalities (Sources F, G, and H) had marginal contributions.

This marginal contribution likely reflects  methodological constraints or a low predictive signal from these sources. Source H’s small sample size (n=70) limits stable estimation of modality-specific weights. Source F’s sparse mutation data likely produces near-constant risk scores, so permutation importance exchanges equivalent values and underestimates its prognostic value. The winner-takes-all feature selection in component-wise boosting further concentrates importance on the strongest modality, masking correlated sources. Redundancy likely explains Source G’s modest importance: when PD-L1 (Source E) and routine inflammatory markers (Source B) are available together, they seem to capture most of the prognostic signal.

\section{Discussion}

We introduced Multimodality Stacking with Blockwise Missing Values (MSB), a late-fusion framework that addresses two coupled challenges in clinical survival analysis: blockwise missing data and high-dimensional source heterogeneity. Both are pronounced in multi-center cohorts such as PIONeeR, where 39\% of entries were missing under source-specific collection protocols~\cite{barlesi_integrative_2026} and where integrating heterogeneous sources drives dimensionality up at the cost of stability~\cite{salerno_high-dimensional_2023}. MSB reduces the meta-learner's input to roughly 24 per-source risk scores while matching or exceeding baselines trained on the full feature space.

Block-aware models outperformed their unified baselines in most families (component-wise boosting being the exception), indicating that the PIONeeR cohort carries exploitable source structure. MSB itself improved on every baseline it wraps (linear $+14.4\%$, forest $+5.0\%$, boosting $+1.8\%$), with the gain inversely related to baseline strength: modality-level decomposition helps most where a singe learner struggles with high-dimensional heterogeneity, and least where the base learner already regularizes heavily. Even component-wise boosting, which performs implicit feature selection, gained from the hierarchical structure ($\Delta$C-index $=+0.012$, Holm-corrected $p=0.037$) ; though this is the smallest and least robust of the four margins.

BlockForest and priorityLasso, the closest architectural counterparts to MSB, were competitive on PIONeeR; the margins were small. Because MSB's base learners are fixed across variants, this spread between parity and advantage is
attributable to the aggregation step alone, making the meta-learner the decisive design choice. MSB's advantage lies less in raw discrimination than in two practical properties: it requires no priority ordering to be fixed in advance
(unlike priorityLasso, whose C-index vary under a permuted order), and it accommodates non-linear aggregation where a linear solution is inadequate ; non-linear meta-learners (RSF, CWXGB) clearly beat linear ones here (Table~\ref{tab:performance}). In this sense MSB extends score-level stacking (StaPLR~\cite{van_loon_imputation_2024}) from linear classification to survival. By design, the per-source risk scores also let the meta-learner predict for block-incomplete patients without listwise deletion ; a property we motivate architecturally here and leave to evaluate directly in future work. Extending BlockForest to test-time missingness via missing-incorporated-in-attributes splits would be a natural forest-native alternative we did not pursue.

This study has several limitations. First, all findings come from a single cohort (PIONeeR, $n=443$, 17 French centers) without external validation; SurvBoard validates the algorithm across 21 public cohorts, but the PIONeeR biomarker findings ; the dominance of Sources A, B, and E, and the near-random discrimination of tumor-derived Sources F, G, and H in isolation ; require prospective replication before clinical interpretation, and may reflect assay or sample-quality constraints in this cohort rather than the underlying biology. Second, we evaluated only two endpoints (PFS and OS); because immuno-oncology involves competing risks and a long-term responder plateau (curves flatten after 1000 days), extending MSB to competing-risk~\cite{alberge_p52_2025} or cure model~\cite{wang_improved_2025} settings is a natural next step. Third, we did not formally characterize the missingness mechanism, using downstream performance as a proxy. Finally, hyperparameters were library defaults on PIONeeR and only lightly adjusted on SurvBoard. Pushing tuning could shift the reported margins, and since MSB variants share the same stacked risk scores, tuning only the meta-learner is a path we could explore.

\section{Conclusion}

This study introduced MSB, a late-fusion stacking framework designed for multi-source survival analysis in high-dimensional datasets with blockwise missingness. By aggregating source-specific predictions into a reduced risk-score space, MSB preserved cohort size without extensive cross-modality imputation, and predicted for patients whose blocks were incomplete at test time. Validation on the PIONeeR dataset showed that MSB improved on the unified baselines it wraps and, with the appropriate meta-learner, exceeded both comparable block-aware algorithms, with the choice of meta-learner emerging as the decisive design factor.

\section{Funding Statement}

This project has received funding from the Excellence Initiative of Aix-Marseille Université - AMidex, a French “Investissements d’Avenir programme” AMX-21-IET-017.

This PIONeeR RHU was funded by a partnership of Aix-Marseille Université (AMU), Assistance Publique Hôpitaux de Marseille (APHM), Centre National de La Recherche Scientifique (CNRS), Institut National de la Santé et de la Recherche Médicale (INSERM), Centre Léon Bérard (CLB), Institut Paoli Calmettes (IPC), Gustave Roussy (GR), AstraZeneca (AZ), Veracyte (VERA), Innate Pharma (IPH) \& ImCheck Therapeutics (ICT), and initiated by Marseille Immunopole.

\section{ Acknowledgment }

We are grateful to Julie Josse for her invaluable guidance on both the methodological framework and the manuscript's structure. Her insights significantly refined the final paper.

\vline 

Implementation of the MSB framework is available \href{https://github.com/MohamedBoussena/MSB}{here} under Inria license.

\newpage

\section*{Appendix}

\renewcommand{\thepage}{S\arabic{page}}
\renewcommand{\thesection}{S\arabic{section}}
\renewcommand{\thetable}{S\arabic{table}}
\renewcommand{\thefigure}{S\arabic{figure}}
\renewcommand{\thealgorithm}{S\arabic{algorithm}}
\renewcommand{\figurename}{Supplementary Figure}
\setcounter{figure}{0}

\begin{figure}[H]
    \centering
    \includegraphics[width=1\linewidth]{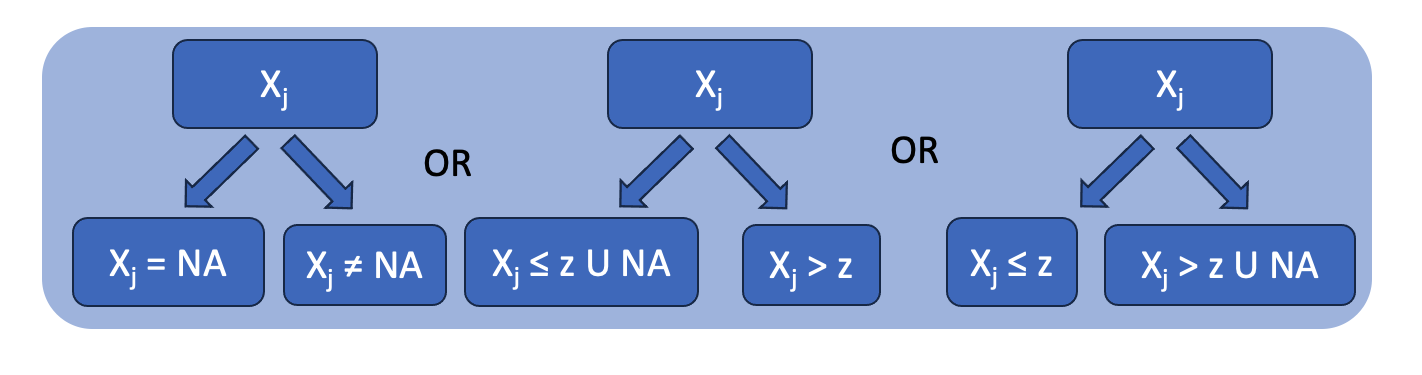}
    \caption{Missing in attributes splitting options, for a given split, of a feature j at value z. Given a splitting criterion, missing values are either sided against values that are not missing or on one side of the splitting value.}
    \label{fig:MIAschema}
\end{figure}

MIA (missing incorporated in attributes) is represented in Figure \ref{fig:MIAschema}.

\begin{figure}[H]
    \centering
    \includegraphics[width=1\linewidth]{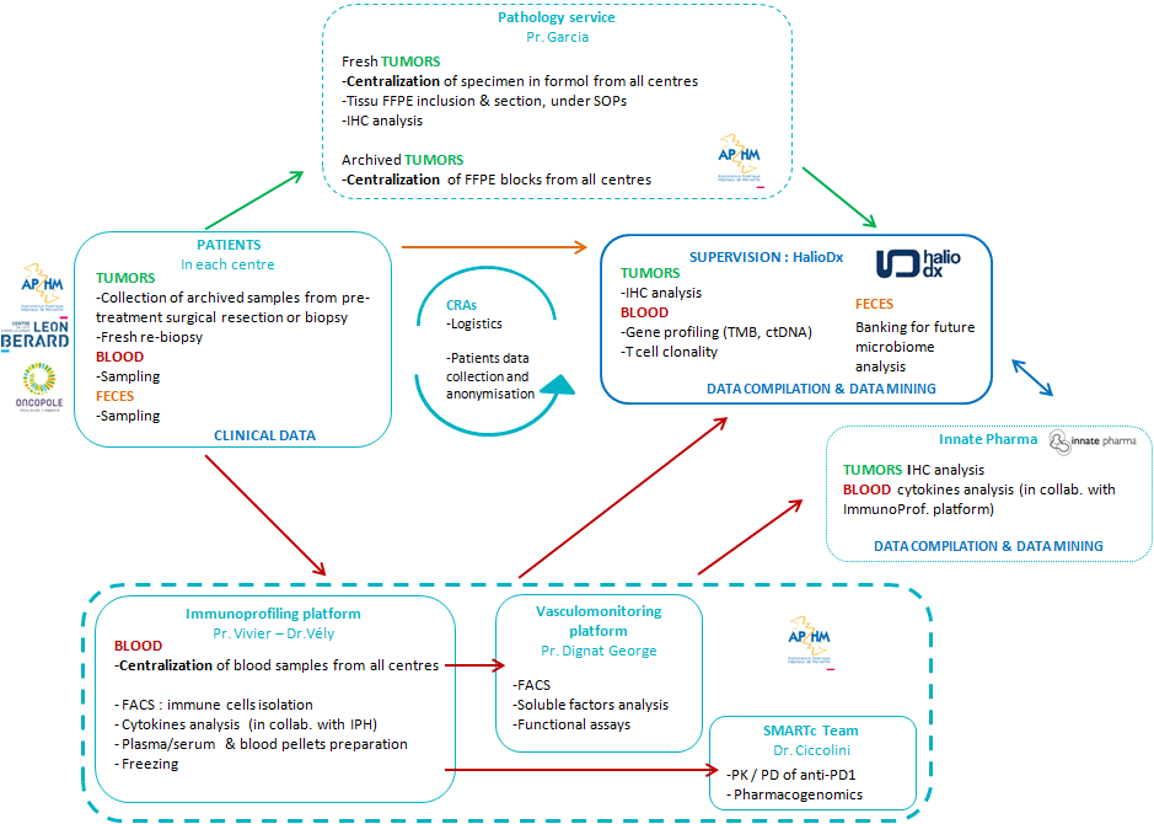}
    \caption{How the dataset is built from the patients to labs.\\ Patients that are part of the study are, for each center, sampled (Tumor and Blood) and described with their clinical biomarkers. The samples are then distributed to each partner responsible of a specific analysis in the study. After this steps, the different analysis results are checked and centralized by Veracyte (formerly HallioDx) for a data lock, ensuring each partner have the same data (the 'Data Compilation' part of the scheme).}
    \label{fig:partner-role}
\end{figure}

\begin{figure}[H]
    \centering
    \includegraphics[width=0.9\linewidth]{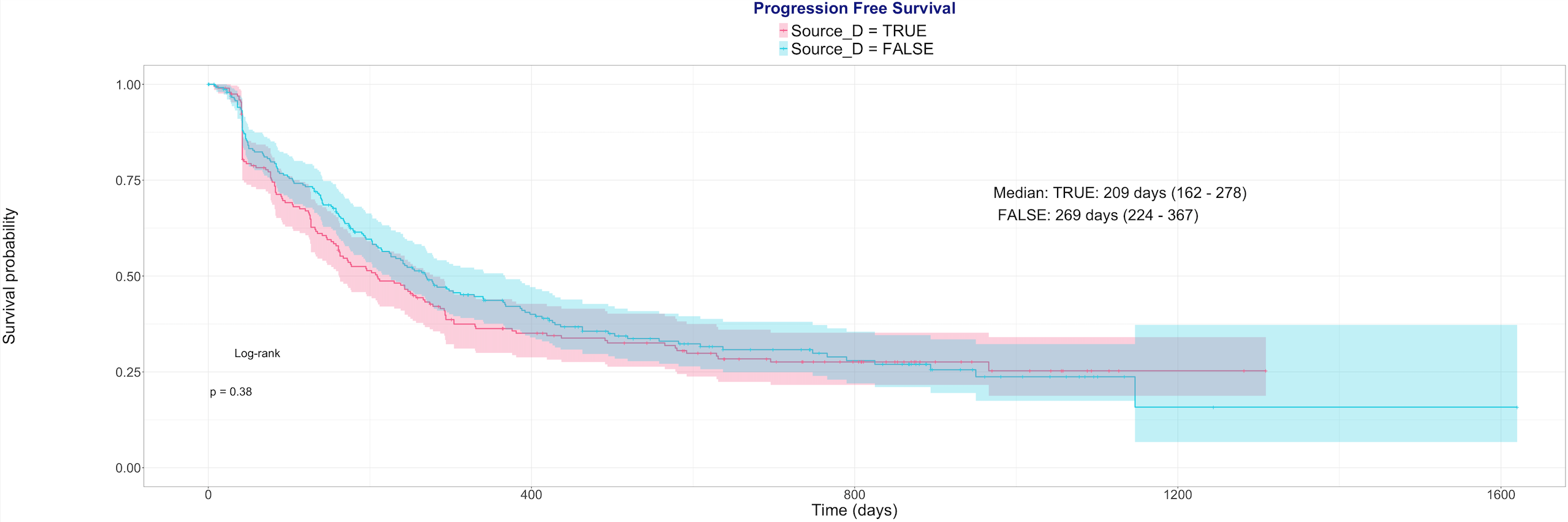}
    \caption{Kaplan–Meier curve of patients grouped according to their originating center. The log-rank test is not significant, indicating no statistically significant difference between the two curves. This suggests that the center of origin does not bias survival.}
    \label{fig:KMPLot}
\end{figure}

\begin{figure}[H]
    \centering
    \includegraphics[width=1\linewidth]{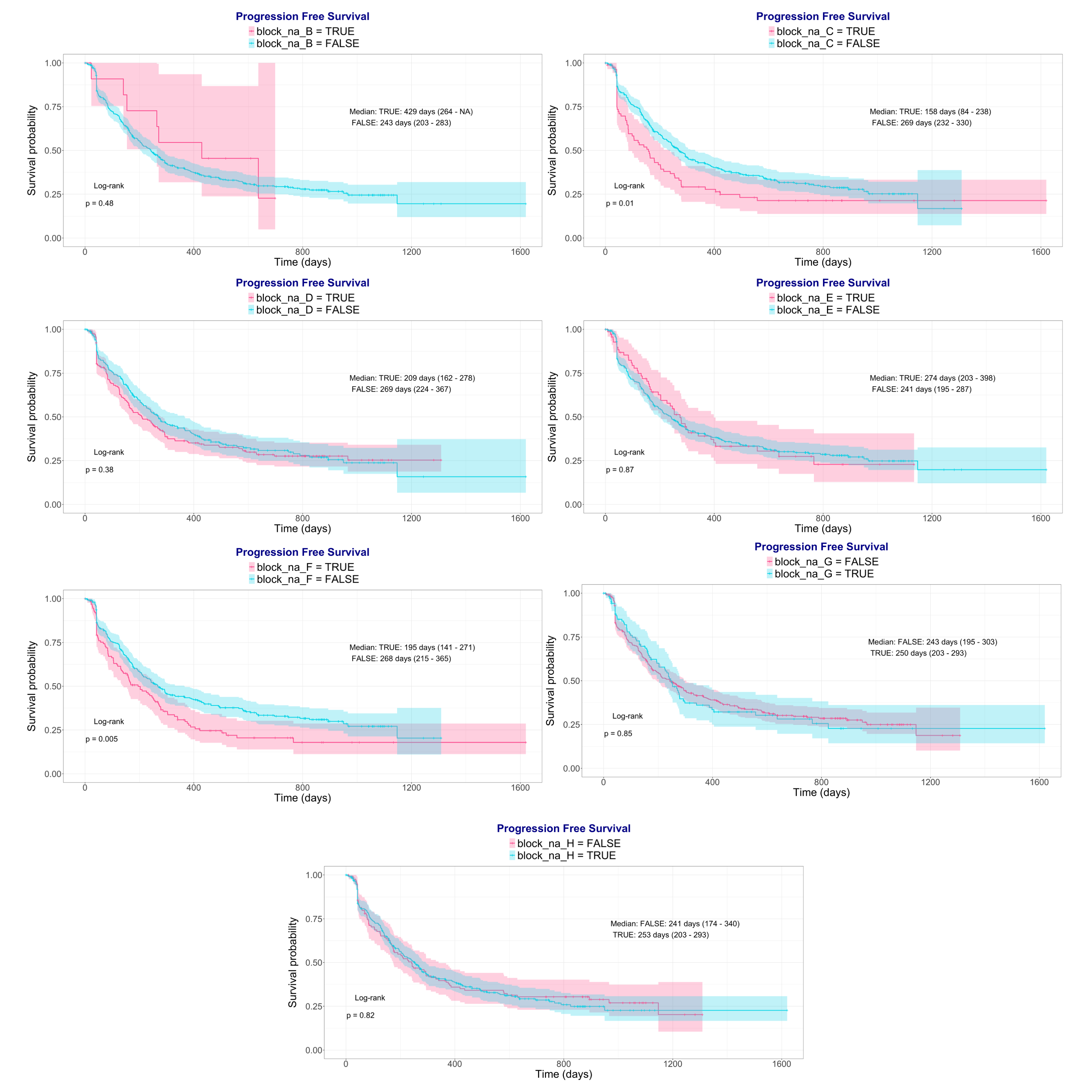}
    \caption{Kaplan-Meier curves stratified by source-level missingness.
Patients were grouped by presence ($\geq 50\%$ complete) vs. absence ($< 50\%$ complete) of each data source. Source A is not shown as it exhibited $<5\%$ missingness. Log-rank p-values are unadjusted for multiple comparisons. Two sources showed nominally significant associations: Source C (immunophenotyping, p=0.01) and Source F (mutations, p=0.005)}
    \label{fig:S4}
\end{figure}

\begin{figure}
    \centering
    \includegraphics[width=0.9\linewidth]{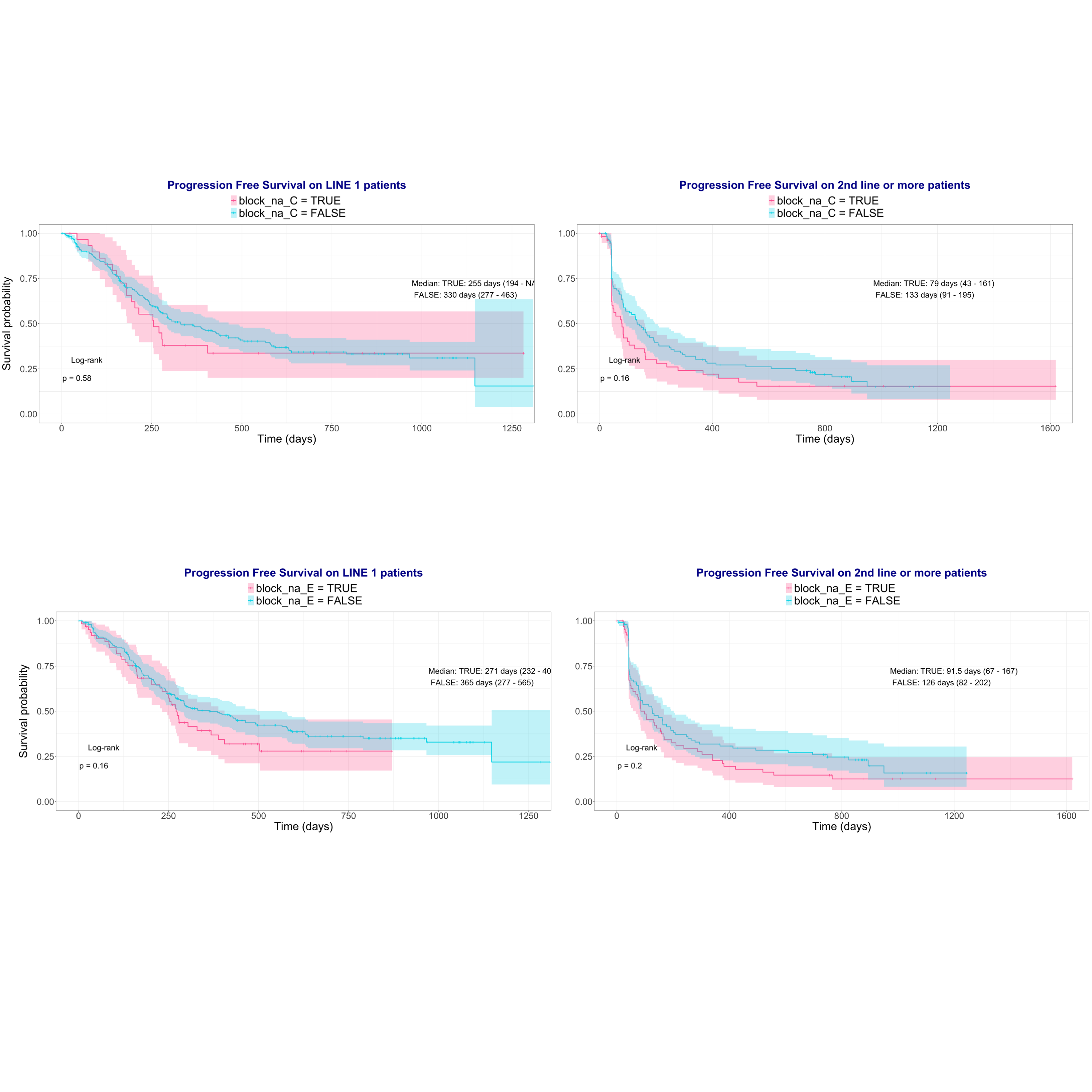}
    \caption{KM plots filtered by LINE of patients on sources C and E. All plots shows no significant survival differences between patients with and without available data.}
    \label{fig:S5}
\end{figure}

To assess whether blockwise missingness was associated to patient prognosis, we generated Kaplan–Meier survival curves stratified by missingness status for each data source (Figure~\ref{fig:S4}). Source A was excluded from this analysis because it had almost complete data ($<5\%$ missingness). For most sources (B, D, E, G, H), log-rank tests indicated no significant survival differences between patients with and without available data (all $p>0.05$). However, two sources showed nominally significant associations: Source C (immunophenotyping, $p=0.01$) and Source F (mutations, $p=0.005$). To determine whether these associations represented independent prognostic effects or confounding by treatment line, we repeated the analyses separately within each treatment line cohort( (Figure~\ref{fig:S5}). Within treatment line strata, the survival differences associated with Source C and F missingness were no longer statistically significant (all $p>0.05$), indicating that treatment line accounted for the observed associations in unadjusted analyses.

\subsection*{MSBs' performance on Death (or Overall Survival)}

\begin{table}[H]
\centering
\caption{Performance table for overall survival (5-fold CV, 5 repetitions). Bold indicates family best; $^*$ indicates global best (Test only). All models include missing block indicators. CWXGB = Component-Wise Gradient Boosting, RSF = Random Survival Forest, XGB = Gradient Boosting, MSB = Multimodality Stacking with Blockwise missing values model.}
\label{tab:performance_DC}
\begin{tabular}{lllrrr}
\toprule
Model Family & Model & Dataset & C-index ($\pm$ SD) & iBS Early & iBS Late \\
\midrule
Linear & MSB-CoxNet$_{imp}$ & test & 0.656\,{\scriptsize$\pm$\,0.045} & 0.059\,{\scriptsize$\pm$\,0.010} & 0.225\,{\scriptsize$\pm$\,0.024} \\
 &  & {\scriptsize train} & {\scriptsize 0.934\,$\pm$\,0.014} & {\scriptsize 0.050\,$\pm$\,0.015} & {\scriptsize 0.081\,$\pm$\,0.011} \\
 & MSB-CoxNet & test & \textbf{0.661\,{\scriptsize$\pm$\,0.040}} & 0.058\,{\scriptsize$\pm$\,0.009} & 0.224\,{\scriptsize$\pm$\,0.021} \\
 &  & {\scriptsize train} & {\scriptsize 0.930\,$\pm$\,0.007} & {\scriptsize 0.055\,$\pm$\,0.011} & {\scriptsize 0.079\,$\pm$\,0.007} \\
 & CoxNet$_{imp}$ & test & 0.567\,{\scriptsize$\pm$\,0.039} & 0.143\,{\scriptsize$\pm$\,0.031} & 0.383\,{\scriptsize$\pm$\,0.042} \\
 &  & {\scriptsize train} & {\scriptsize 0.971\,$\pm$\,0.007} & {\scriptsize 0.009\,$\pm$\,0.002} & {\scriptsize 0.032\,$\pm$\,0.005} \\
 & priorityLasso & test & 0.659\,{\scriptsize$\pm$\,0.043} & \textbf{0.055\,{\scriptsize$\pm$\,0.013}} & \textbf{0.203\,{\scriptsize$\pm$\,0.016}} \\
 &  & {\scriptsize train} & {\scriptsize 0.713\,$\pm$\,0.025} & {\scriptsize 0.051\,$\pm$\,0.003} & {\scriptsize 0.180\,$\pm$\,0.009} \\
\midrule
Forest & MSB-RSF & test & 0.673\,{\scriptsize$\pm$\,0.035} & 0.056\,{\scriptsize$\pm$\,0.011} & \textbf{0.202\,{\scriptsize$\pm$\,0.014}} \\
 &  & {\scriptsize train} & {\scriptsize 0.877\,$\pm$\,0.014} & {\scriptsize 0.040\,$\pm$\,0.002} & {\scriptsize 0.115\,$\pm$\,0.009} \\
 & MSB-RSF$_{imp}$ & test & 0.672\,{\scriptsize$\pm$\,0.039} & 0.056\,{\scriptsize$\pm$\,0.011} & 0.203\,{\scriptsize$\pm$\,0.016} \\
 &  & {\scriptsize train} & {\scriptsize 0.858\,$\pm$\,0.021} & {\scriptsize 0.043\,$\pm$\,0.002} & {\scriptsize 0.137\,$\pm$\,0.015} \\
 & MSB-RSF$_{MIA}$ & test & 0.651\,{\scriptsize$\pm$\,0.042} & 0.058\,{\scriptsize$\pm$\,0.010} & 0.212\,{\scriptsize$\pm$\,0.016} \\
 &  & {\scriptsize train} & {\scriptsize 0.884\,$\pm$\,0.015} & {\scriptsize 0.044\,$\pm$\,0.002} & {\scriptsize 0.114\,$\pm$\,0.013} \\
 & RSF$_{imp}$ & test & 0.649\,{\scriptsize$\pm$\,0.043} & \textbf{0.055\,{\scriptsize$\pm$\,0.011}} & 0.206\,{\scriptsize$\pm$\,0.010} \\
 &  & {\scriptsize train} & {\scriptsize 0.941\,$\pm$\,0.003} & {\scriptsize 0.027\,$\pm$\,0.001} & {\scriptsize 0.056\,$\pm$\,0.001} \\
 & RSF$_{MIA}$ & test & 0.644\,{\scriptsize$\pm$\,0.044} & 0.057\,{\scriptsize$\pm$\,0.012} & 0.207\,{\scriptsize$\pm$\,0.011} \\
 &  & {\scriptsize train} & {\scriptsize 0.922\,$\pm$\,0.005} & {\scriptsize 0.033\,$\pm$\,0.002} & {\scriptsize 0.087\,$\pm$\,0.003} \\
 & BlockForest & test & \textbf{0.680\,{\scriptsize$\pm$\,0.037}} & 0.055\,{\scriptsize$\pm$\,0.011} & 0.203\,{\scriptsize$\pm$\,0.010} \\
 &  & {\scriptsize train} & {\scriptsize 0.927\,$\pm$\,0.004} & {\scriptsize 0.028\,$\pm$\,0.001} & {\scriptsize 0.095\,$\pm$\,0.002} \\
\midrule
Boosting & MSB-CWXGB & test & 0.682\,{\scriptsize$\pm$\,0.037} & 0.054\,{\scriptsize$\pm$\,0.012} & 0.198\,{\scriptsize$\pm$\,0.012} \\
 &  & {\scriptsize train} & {\scriptsize 0.754\,$\pm$\,0.018} & {\scriptsize 0.050\,$\pm$\,0.003} & {\scriptsize 0.173\,$\pm$\,0.006} \\
 & MSB-CWXGB$_{imp}$ & test & \textbf{0.683\,{\scriptsize$\pm$\,0.043}$^*$} & \textbf{0.054\,{\scriptsize$\pm$\,0.012}$^*$} & \textbf{0.198\,{\scriptsize$\pm$\,0.014}$^*$} \\
 &  & {\scriptsize train} & {\scriptsize 0.745\,$\pm$\,0.014} & {\scriptsize 0.050\,$\pm$\,0.003} & {\scriptsize 0.175\,$\pm$\,0.005} \\
 & CWXGB$_{imp}$ & test & 0.659\,{\scriptsize$\pm$\,0.036} & 0.055\,{\scriptsize$\pm$\,0.012} & 0.203\,{\scriptsize$\pm$\,0.011} \\
 &  & {\scriptsize train} & {\scriptsize 0.738\,$\pm$\,0.010} & {\scriptsize 0.051\,$\pm$\,0.003} & {\scriptsize 0.178\,$\pm$\,0.003} \\
 & MSB-XGB & test & 0.665\,{\scriptsize$\pm$\,0.040} & 0.057\,{\scriptsize$\pm$\,0.012} & 0.217\,{\scriptsize$\pm$\,0.022} \\
 &  & {\scriptsize train} & {\scriptsize 0.837\,$\pm$\,0.032} & {\scriptsize 0.045\,$\pm$\,0.005} & {\scriptsize 0.139\,$\pm$\,0.021} \\
 & MSB-XGB$_{imp}$ & test & 0.657\,{\scriptsize$\pm$\,0.046} & 0.059\,{\scriptsize$\pm$\,0.012} & 0.222\,{\scriptsize$\pm$\,0.022} \\
 &  & {\scriptsize train} & {\scriptsize 0.790\,$\pm$\,0.066} & {\scriptsize 0.048\,$\pm$\,0.005} & {\scriptsize 0.165\,$\pm$\,0.038} \\
 & XGB$_{imp}$ & test & 0.657\,{\scriptsize$\pm$\,0.048} & 0.056\,{\scriptsize$\pm$\,0.012} & 0.218\,{\scriptsize$\pm$\,0.022} \\
 &  & {\scriptsize train} & {\scriptsize 0.935\,$\pm$\,0.005} & {\scriptsize 0.029\,$\pm$\,0.003} & {\scriptsize 0.075\,$\pm$\,0.004} \\
\bottomrule
\end{tabular}
\end{table}

\subsection*{Hyperparameters for results}

Each Meta learners used in MSB (Coxnet, RSF, CWXGB, XGB)  uses default parameters from scikit survival's API.

For MSB, the stacking procedure uses 5 folds cross validation. the base learners are used with default hyperparameters, as well as the meta-learner.

\subsection*{Modality importances for other MSB models}

\begin{figure}[H]
    \centering
    \includegraphics[width=0.7\linewidth]{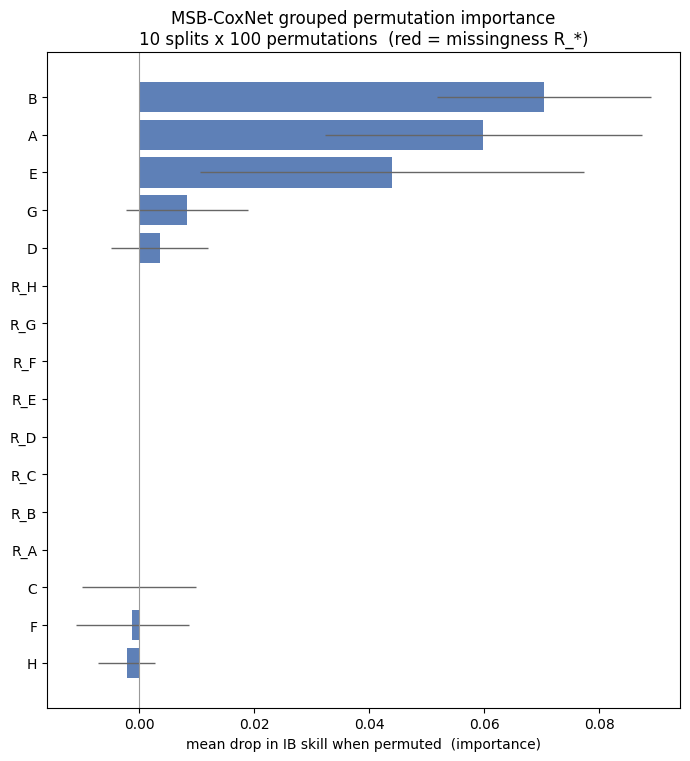}
    \caption{Permutation importance on iBSS for CoxNet as meta model for MSB. }
    \label{fig:imp_PFS_CoxNet}
\end{figure}


 \begin{figure}[H]
    \centering
    \includegraphics[width=0.7\linewidth]{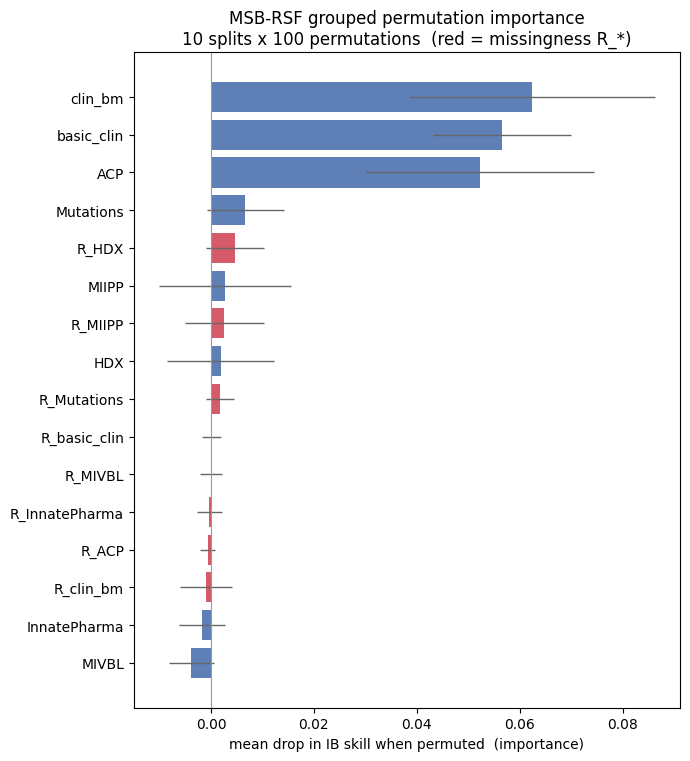}
    \caption{Permutation importance on iBSS for RSF as meta model in MSB}
    \label{fig:imp_PFS_RSF}

\end{figure}

\subsection*{Rank in survboard}

\begin{table}[H]
\caption{Rank by average metric across cohorts}
\label{tab:survboard_rank}
\begin{tabular}{lrrrr}
\toprule
 & rank\_cindex & avg\_antolini & rank\_ibs & avg\_ibs \\
model &  &  &  &  \\
\midrule
MSB (with missing) & 1 & 0.660 & 1 & 0.184 \\
MSB & 2 & 0.627 & 5 & 0.191 \\
priority\_elastic\_net & 3 & 0.626 & 7 & 0.194 \\
priority\_elastic\_net\_missing & 4 & 0.619 & 2 & 0.186 \\
eh\_intermediate\_concat\_missing & 5 & 0.618 & 6 & 0.191 \\
cox\_intermediate\_concat\_missing & 6 & 0.611 & 10 & 0.197 \\
salmon & 7 & 0.609 & 4 & 0.188 \\
blockforest & 8 & 0.608 & 3 & 0.187 \\
eh\_late\_mean & 9 & 0.601 & 12 & 0.199 \\
cox\_late\_mean & 10 & 0.599 & 14 & 0.203 \\
customics & 11 & 0.599 & 11 & 0.198 \\
multimodal\_survival\_pred & 12 & 0.596 & 17 & 0.269 \\
eh\_intermediate\_concat & 13 & 0.592 & 9 & 0.196 \\
cox\_intermediate\_concat & 14 & 0.590 & 13 & 0.201 \\
multimodal\_nsclc & 15 & 0.589 & 15 & 0.220 \\
gdp & 16 & 0.581 & 8 & 0.194 \\
survival\_net & 17 & 0.570 & 16 & 0.222 \\
\bottomrule
\end{tabular}
\end{table}

(for reference, on MSB (with missing)) Friedman c-index: $chi2=112.2$, $p=1.7e-16$

(for reference, on MSB (with missing)) Friedman IBS    : $chi2=158.9$, $p=1.4e-25$

\subsection*{Survboard MSB benchmark models hyperparameter and results on other models}

\begin{table}[H]
\centering
\caption{Hyperparameter configuration for the MSB late-fusion estimator on the full
21-cancer TCGA benchmark (complete-modality \emph{full} setting, official 25-fold splits).
Any parameter not listed is left at its \texttt{scikit-survival}~0.27 / \texttt{scikit-learn} default. The reported model uses the ComponentwiseGradientBoosting (CWGB) meta-learner; RSF and CoxNet meta-learners are reported as ablations.}

\label{tab:msb-hyperparams}
\small
\begin{tabular}{@{}lll@{}}
\toprule
Stage & Component & Setting \\
\midrule
\multicolumn{3}{@{}l}{\emph{Per-block preprocessing}}\\
\addlinespace[2pt]
Imputation      & sporadic missing values      & median (per feature) \\
Standardization & all blocks                   & zero mean, unit variance \\
Dim.\ reduction & gex, cnv, meth, mut ($>$20k feat.) & PCA, $k=20$ components \\
                & clinical, mirna, rppa        & kept raw (no PCA) \\
\midrule
\multicolumn{3}{@{}l}{\emph{Base (source) learners ; one set per block, out-of-fold risk scores}}\\
\addlinespace[2pt]
Learner 1 & Random Survival Forest & $n_{\text{trees}}=100$, min.\ leaf $=15$ \\
Learner 2 & Componentwise Grad.\ Boosting & $n_{\text{est}}=200$, learning rate $=0.1$ \\
Stacking CV & internal out-of-fold split & 5 folds \\
Base output & score passed to meta-learner & linear predictor (risk) \\
\midrule
\multicolumn{3}{@{}l}{\emph{Meta (stacking) learner ; on stacked block risk scores}}\\
\addlinespace[2pt]
Primary   & Componentwise Grad.\ Boosting & $n_{\text{est}}=200$, learning rate $=0.1$ \\
Ablation  & Random Survival Forest        & $n_{\text{trees}}=100$, max depth $=5$ \\
Ablation  & CoxNet (elastic net Cox)      & $\ell_1$ ratio $=0.5$, standardized input \\
\midrule
\multicolumn{3}{@{}l}{\emph{Fusion / evaluation protocol}}\\
\addlinespace[2pt]
Missing-block handling & complete-modality setting & disabled (no indicators / imputation) \\
Cross-validation       & SurvBoard official splits & 25 folds \\
Discrimination         & concordance               & Antolini (pycox \texttt{adj\_antolini}) \\
Calibration            & integrated Brier score    & grid $linspace(t_{\min}^{\text{test}}, t_{\max}^{\text{test}}, 100)$, \\
                       &                           & \phantom{grid }test-set IPCW censoring \\
\bottomrule
\end{tabular}
\end{table}

\section{Model configurations and hyperparameters}
\label{app:hyperparameters}

All learners were used through their reference implementations: the Python survival
models via \texttt{scikit-survival} 0.27, and \texttt{blockForest} via the original
R package \cite{hornung_block_2019}. Hyperparameters not stated below were left
at their library defaults; random seeds were fixed for reproducibility.

\subsection{Evaluation protocol}
\label{app:protocol}

Models were compared under repeated stratified $k$-fold cross-validation
(\texttt{RepeatedStratifiedKFold}), $k=5$ folds $\times\,5$ repeats (25 train/test
folds), stratified on the event indicator, with a fixed seed. All preprocessing
(imputation, block-wise fitting) was performed on the training fold only and applied
to the held-out fold. Discrimination was measured with Harrell's concordance index
(and, where noted, Antolini's time-dependent concordance); calibration with the
integrated Brier score (IBS) computed on two follow-up windows, \emph{early}
$[10,100]$ and \emph{late} $[100,1000]$ days, on a 50-point time grid, using the
held-out set's Kaplan--Meier estimate as the censoring distribution.

\subsection{Imputation}
\label{app:imputation}

Two imputation regimes were used. (i) \emph{Global} imputation with a $k$-nearest-neighbours
imputer (\texttt{KNNImputer}, $k=5$ neighbours, uniform weights), fit on the training
fold, used for the single-model early-fusion baselines and for the ``$_{\mathrm{imp}}$''
MSB variants. (ii) \emph{Native} block-wise handling inside MSB (Sec.~\ref{app:msb}),
where each modality block is imputed independently with a mean imputer and a
block-missingness indicator is retained. The R \texttt{blockForest} baseline, which has
no native missing-value handling, was fed the globally KNN-imputed data.

\subsection{Multimodality Stacking with Block-wise missingness (MSB)}
\label{app:msb}

MSB is a late-fusion stacking estimator. For each of the $M$ modality blocks, a fixed
set of base learners produces an out-of-fold risk score via an internal
$5$-fold cross-validation; the resulting per-block scores, together with per-block
missingness indicators, form the meta-features passed to a final estimator.

\begin{table}[htbp]
\centering
\caption{MSB architecture and hyperparameters.}
\label{tab:msb-hyper}
\begin{tabular}{ll}
\toprule
Component & Setting \\
\midrule
Modality blocks ($M$)        & 8 (\texttt{basic\_clin, clin\_bm, HDX, MIVBL,} \\
                             & \texttt{MIIPP, ACP, Mutations, InnatePharma}) \\
Base learners (per block)    & CWGB + RSF (fixed; see Tab.~\ref{tab:msb-hyperparams}) \\
Final / meta estimator       & varied: CoxNet, RSF, CWGB, or XGB \\
Internal CV (out-of-fold)    & 5-fold \\
Meta-features                & per-block risk scores $+$ per-block missingness ratios \\
Block-output summary         & base-learner risk score \\
Block inclusion rule         & a block's base learners are trained on samples with \\
                             & $<50\%$ missing entries in that block \\
Meta-feature imputation      & mean imputation (\texttt{SimpleImputer}) of absent \\
                             & block scores before the final estimator \\
Passthrough (raw) features   & none \\
\bottomrule
\end{tabular}
\end{table}

Two configurations of each MSB variant are reported. In the \emph{native} configuration
(e.g.\ \textsc{MSB-CWGB}), the raw data with missing values is passed to MSB, which
performs block-wise KNN imputation and retains the missingness indicators. In the
\emph{imputed} configuration (``$_{\mathrm{imp}}$''), the dataset is globally KNN-imputed
beforehand, so MSB reduces to pure late fusion and the missingness indicators are
constant. Only the final estimator varies across MSB variants; the per-block base
learners are held fixed.

\subsection{Base and baseline learners}
\label{app:baselearners}

\begin{table}[htbp]
\centering
\caption{Random Survival Forest (RSF; \texttt{RandomSurvivalForest}).}
\label{tab:rsf-hyper}
\begin{tabular}{ll}
\toprule
Hyperparameter & Value \\
\midrule
Number of trees                  & 100 (default) \\
Minimum samples per leaf         & 3 (default) \\
Max.\ features per split         & $\sqrt{p}$ (default) \\
Bootstrap                        & yes (default) \\
Random state                     & fixed \\
\bottomrule
\end{tabular}
\end{table}

\subsection{BlockForest (R baseline)}
\label{app:blockforest}

The block-structured random forest was run through
the original R \texttt{blockForest} package (function \texttt{blockfor}), on the
globally KNN-imputed data. Covariates were partitioned into the same $M$ modality
blocks; where a feature was shared between blocks it was assigned to the first block.
Risk scores were taken as the summed predicted cumulative hazard, and survival curves
from the underlying \texttt{ranger} ensemble.

\begin{table}[htbp]
\centering
\caption{BlockForest (\texttt{blockForest::blockfor}).}
\label{tab:blockforest-hyper}
\begin{tabular}{ll}
\toprule
Hyperparameter & Value \\
\midrule
Block method                 & \texttt{BlockForest} \\
Number of trees (final)      & 100 \\
Tuning iterations (\texttt{nsets})       & 20 \\
Trees per pre-tuning forest (\texttt{num.trees.pre}) & 50 \\
Split rule                   & \texttt{extratrees} (default) \\
Minimum node size            & 3 (survival default) \\
Block partition              & 8 modality blocks (first-block-wins on overlap) \\
Random state                 & fixed \\
\bottomrule
\end{tabular}
\end{table}

\noindent\emph{Note.} As an MSB \emph{base} learner, RSF used the library default

\newpage
\printbibliography

\end{document}